\PassOptionsToPackage{unicode}{hyperref}
\PassOptionsToPackage{hyphens}{url}
\documentclass[
  12pt,
]{article}
\usepackage{amsmath,amssymb}
\usepackage{iftex}
\ifPDFTeX
  \usepackage[T1]{fontenc}
  \usepackage[utf8]{inputenc}
  \usepackage{textcomp} 
\else 
  \usepackage{unicode-math} 
  \defaultfontfeatures{Scale=MatchLowercase}
  \defaultfontfeatures[\rmfamily]{Ligatures=TeX,Scale=1}
\fi
\usepackage{lmodern}
\ifPDFTeX\else
\fi
\IfFileExists{upquote.sty}{\usepackage{upquote}}{}
\IfFileExists{microtype.sty}{
  \usepackage[]{microtype}
  \UseMicrotypeSet[protrusion]{basicmath} 
}{}
\makeatletter
\@ifundefined{KOMAClassName}{
  \IfFileExists{parskip.sty}{%
    \usepackage{parskip}
  }{
    \setlength{\parindent}{0pt}
    \setlength{\parskip}{6pt plus 2pt minus 1pt}}
}{
  \KOMAoptions{parskip=half}}
\makeatother
\usepackage{xcolor}
\usepackage[margin=1in]{geometry}
\usepackage{color}
\usepackage{fancyvrb}

\DefineVerbatimEnvironment{Highlighting}{Verbatim}{commandchars=\\\{\}}
\usepackage{framed}
\definecolor{shadecolor}{RGB}{248,248,248}
\newenvironment{Shaded}{\begin{snugshade}}{\end{snugshade}}

\newcommand{\AttributeTok}[1]{\textcolor[rgb]{0.13,0.29,0.53}{#1}}

\newcommand{\CommentTok}[1]{\textcolor[rgb]{0.56,0.35,0.01}{\textit{#1}}}

\newcommand{\ControlFlowTok}[1]{\textcolor[rgb]{0.13,0.29,0.53}{\textbf{#1}}}
\newcommand{\DataTypeTok}[1]{\textcolor[rgb]{0.13,0.29,0.53}{#1}}
\newcommand{\DecValTok}[1]{\textcolor[rgb]{0.00,0.00,0.81}{#1}}

\newcommand{\FloatTok}[1]{\textcolor[rgb]{0.00,0.00,0.81}{#1}}
\newcommand{\FunctionTok}[1]{\textcolor[rgb]{0.13,0.29,0.53}{\textbf{#1}}}
\newcommand{\ImportTok}[1]{#1}

\newcommand{\KeywordTok}[1]{\textcolor[rgb]{0.13,0.29,0.53}{\textbf{#1}}}
\newcommand{\NormalTok}[1]{#1}
\newcommand{\OperatorTok}[1]{\textcolor[rgb]{0.81,0.36,0.00}{\textbf{#1}}}
\newcommand{\OtherTok}[1]{\textcolor[rgb]{0.56,0.35,0.01}{#1}}
\newcommand{\PreprocessorTok}[1]{\textcolor[rgb]{0.56,0.35,0.01}{\textit{#1}}}

\newcommand{\SpecialCharTok}[1]{\textcolor[rgb]{0.81,0.36,0.00}{\textbf{#1}}}

\newcommand{\StringTok}[1]{\textcolor[rgb]{0.31,0.60,0.02}{#1}}

\usepackage{longtable,booktabs,array}
\usepackage{calc} 
\usepackage{etoolbox}
\makeatletter
\patchcmd\longtable{\par}{\if@noskipsec\mbox{}\fi\par}{}{}
\makeatother
\IfFileExists{footnotehyper.sty}{\usepackage{footnotehyper}}{\usepackage{footnote}}
\makesavenoteenv{longtable}
\usepackage{graphicx}
\makeatletter
\newsavebox\pandoc@box
\newcommand*\pandocbounded[1]{
  \sbox\pandoc@box{#1}%
  \Gscale@div\@tempa{\textheight}{\dimexpr\ht\pandoc@box+\dp\pandoc@box\relax}%
  \Gscale@div\@tempb{\linewidth}{\wd\pandoc@box}%
  \ifdim\@tempb\p@<\@tempa\p@\let\@tempa\@tempb\fi
  \ifdim\@tempa\p@<\p@\scalebox{\@tempa}{\usebox\pandoc@box}%
  \else\usebox{\pandoc@box}%
  \fi%
}
\def\fps@figure{htbp}
\makeatother
\NewDocumentCommand\citeproctext{}{}

\makeatletter
 \let\@cite@ofmt\@firstofone
 \def\@biblabel#1{}
 \def\@cite#1#2{{#1\if@tempswa , #2\fi}}
\makeatother
\newlength{\cslhangindent}
\newlength{\csllabelwidth}
\newenvironment{CSLReferences}[2] 
 {\begin{list}{}{%
  \setlength{\itemindent}{0pt}
  \setlength{\leftmargin}{0pt}
  \setlength{\parsep}{0pt}
  \ifodd #1
   \setlength{\leftmargin}{\cslhangindent}
   \setlength{\itemindent}{-1\cslhangindent}
  \fi
  \setlength{\itemsep}{#2\baselineskip}}}
 {\end{list}}
\usepackage{calc}

\usepackage{tcolorbox}
\usepackage{amssymb}
\usepackage{yfonts}
\usepackage{bm}
\usepackage{titlesec}

\newtcolorbox{greybox}{
  colback=white,
  colframe=blue,
  coltext=black,
  boxsep=5pt,
  arc=4pt}

\usepackage{bookmark}
\IfFileExists{xurl.sty}{\usepackage{xurl}}{} 
\hypersetup{
  pdftitle={Higher Partials of fStress},
  pdfauthor={Jan de Leeuw - University of California Los Angeles},
  hidelinks,
  pdfcreator={LaTeX via pandoc}}

\title{Higher Partials of fStress}
\author{Jan de Leeuw - University of California Los Angeles}
\date{Started July 20 2017, Version of July 25, 2024}

\begin{document}
\maketitle
\begin{abstract}
We define \emph{fDistances}, which generalize Euclidean distances, squared distances, and log distances. The least squares loss function to fit fDistances to dissimilarity data is \emph{fStress}. We give formulas and R/C code to compute partial derivatives of orders one to four of fStress, relying heavily on the use of Faà di Bruno's chain rule formula for higher derivatives. ---
\end{abstract}

{
\setcounter{tocdepth}{3}
\tableofcontents
}
Note: This is a working paper which will be expanded/updated frequently. All suggestions for improvement are welcome. The directory \href{http://deleeuwpdx.net/pubfolders/fStress}{deleeuwpdx.net/pubfolders/fStress} has a pdf version, the bib file, the complete Rmd file with the code chunks, and the R and C source code.

\section{Introduction}\label{introduction}

The multidimensional scaling (MDS) loss function \emph{fStress} (Groenen, De Leeuw, and Mathar (1995)) is defined as
\begin{equation}\label{E:fstress}
\sigma(x):=\mathop{\sum\sum}_{1\leq i<j\leq n}w_{ij}(\delta_{ij}-f(x'A_{ij}x))^2
\end{equation}
for some real-valued \(f\). The \(w_{ij}\) are positive weights, the \(\delta_{ij}\) are \emph{dissimilarities}. This does not necessarily imply that they are actual dissimilarity judgments or measurements, they can be arbitrary transformations of such judgments or measurements as well. Both \(w_{ij}\) and \(\delta_{ij}\) are fixed and known numbers, the \(A_{ij}\) are fixed and known symmetric matrices. Note there is no constraint that either the dissimilarities \(\delta_{ij}\) or the fDistances \(f(x'A_{ij}x)\) are non-negative, or that \(f\) is increasing. Thus minimizing fStress can also be used, for example, to fit fDistances to similarities.

Our notation is somewhat different from the standard MDS notation, so let's make some translations. We use \(x=\mathbf{vec}(X)\), with \(X\) the MDS configuration of \(n\) points in \(p\) dimensions. If we define \(\nu(i,s)=(s-1)*n+i\), then
\(\{X\}_{is}=x_{\nu(i,s)}\). For most of our formulas it is easier to work with vectors than it is to work with matrices, so we will use
\(x\) instead of \(X\). It is also helpful for our software development, since matrices and higher-dimensional arrays are stored as
vectors anyway.

The \(np\times np\) matrices \(A_{ij}\) are defined using unit vectors \(e_i\) and \(e_j\) of length \(n\), all zero except for one element that is equal to one. If you like, the \(e_i\) are the columns of the identity matrix. We first define \(n\times n\) matrices \(\mathcal{A}_{ij}\) by
\[
\mathcal{A}_{ij}:=(e_i-e_j)(e_i-e_j)',
\]
and use \(p\) copies of \(\mathcal{A}_{ij}\) to make the direct sum
\[
A_{ij}:=\underbrace{\mathcal{A}_{ij}\oplus\cdots\oplus\mathcal{A}_{ij}}_{p\text{ times}}.
\]
Thus the squared Euclidean distance between point \(i\) and \(j\) in the configuration \(X\) is \(\mathbf{tr}\ X'\mathcal{A}_{ij}X=x'A_{ij}x\).

\subsection{fDistances}\label{fdistances}

An \emph{fDistance} is a function of the form \(g(x)=f(x'Ax)\) for some real-valued \(f\). Thus minimzinf fStress is fitting fDistances to dissimilarities, using weighted least squares. fStress properly generalizes the usual \emph{stress} function initially proposed by Kruskal (1964a) and Kruskal (1964b), which uses \(g(x)=\sqrt{x'Ax}\). The \emph{sStress} loss function of Takane, Young, and De Leeuw (1977) uses the identity \(g(x)=x'Ax\), and the \emph{lStress} function of Ramsay (1977) and Ramsay (1982) uses the logarithm \(g(x)=\log(x'Ax)\).

For \(g\) a general power, i.e.~\(g(x)=(x'Ax)^r\), we get \emph{rStress}, sometimes also known as \emph{qStress}, studied in a host of (unpublished, my fault) papers by De Leeuw, Groenen, and Pietersz (2006), Groenen and De Leeuw (2010), De Leeuw (2014), De Leeuw, Groenen, and Mair (2016c), De Leeuw, Groenen, and Mair (2016a), De Leeuw, Groenen, and Mair (2016d), De Leeuw, Groenen, and Mair (2016b). URL's and/or DOI's are in the references section.

In Groenen, De Leeuw, and Mathar (1995) first and second partials of fStress are given. We add third and fourth order partials. It is unclear if the higher order partials have any practical applications. In De Leeuw, Groenen, and Mair (2016d) there are some applications of the second derivatives of rStress. We briefly mention some possible applications of our higher order partials to majorization (a.k.a. MM) algorithms.

\section{Partials, Partials Everywhere}\label{partials-partials-everywhere}

\subsection{Derivatives of fDistances}\label{derivatives-of-fdistances}

We give expressions for the first four derivatives of an arbitrary, but sufficiently many times differentiable, fDistance. In fact, we first address a more general problem, which in its turn is a special case of the first four terms of the multivariate Faà di Bruno formula (see, for instance, Constantine and Savits (1996), Leipnik and Pearce (2007)). We then apply those results to fDistances.

\subsection{Faà di Bruno}\label{fauxe0-di-bruno}

Suppose \(h:\mathcal{R}^n\rightarrow\mathcal{R}\), \(g:\mathcal{R}^n\rightarrow\mathcal{R}\), and \(f:\mathcal{R}\rightarrow\mathcal{R}\), such that \(h(x)=f(g(x))\). We will give expressions for the partials of \(h\) of orders up to four. Our formulas separate the parts that depend on \(f\) from the parts that depend only on \(g\).

\subsubsection{First}\label{first}

The first partials can be written in the form
\[
\mathcal{D}_ih(x)=\mathcal{D}f(g(x))h_{11}(x),
\]
with
\[
h_{11}(x):=\mathcal{D}_ig(x).
\]

\subsubsection{Second}\label{second}

Next
\[
\mathcal{D}_{ij}h(x)=\mathcal{D}f(g(x))h_{21}(x)+\mathcal{D}^2f(g(x))h_{22}(x),
\]
\begin{align*}
h_{21}(x)&:=\mathcal{D}_{ij}g(x),\\
h_{22}(x)&:=\mathcal{D}_ig(x)\mathcal{D}_jg(x).
\end{align*}

\subsubsection{Third}\label{third}

Next
\[
\mathcal{D}_{ijk}h(x)=\mathcal{D}f(g(x))h_{31}(x)+\mathcal{D}^2f(g(x))h_{32}(x)+\mathcal{D}^3f(g(x))h_{33}(x),
\]
with
\begin{align*}
h_{31}(x)&:=\mathcal{D}_{ijk}g(x),\\
h_{32}(x)&:=\mathcal{D}_jg(x)\mathcal{D}_{ik}g(x)+\mathcal{D}_ig(x)\mathcal{D}_{jk}g(x)+\mathcal{D}_kg(x)\mathcal{D}_{ij}g(x),\\
h_{33}(x)&:=\mathcal{D}_ig(x)\mathcal{D}_jg(x)\mathcal{D}g_k(x).
\end{align*}

\subsubsection{Fourth}\label{fourth}

And finally
\[
\mathcal{D}_{ijkl}h(x)=\mathcal{D}f(g(x))h_{41}(x)+\mathcal{D}^2f(g(x))h_{42}(x)+\mathcal{D}^3f(g(x))h_{43}(x)+\mathcal{D}^4f(g(x))h_{44}(x),
\]
with
\begin{align*}
h_{41}(x)&:=\mathcal{D}_{ijkl}g(x),\\
h_{42}(x)&:=\mathcal{D}_{jl}g(x)\mathcal{D}_{ik}g(x)+\mathcal{D}_{il}g(x)\mathcal{D}_{jk}g(x)\\
&+\mathcal{D}_jg(x)\mathcal{D}_{ikl}g(x)+\mathcal{D}_ig(x)\mathcal{D}_{jkl}g(x)+\mathcal{D}_kg(x)\mathcal{D}_{ijl}g(x)+\mathcal{D}_lg(x)\mathcal{D}_{ijk}g(x),\\
h_{43}(x)&:=\mathcal{D}_{il}g(x)\mathcal{D}_jg(x)\mathcal{D}g_k(x)+\mathcal{D}_ig(x)\mathcal{D}_{jl}g(x)\mathcal{D}g_k(x)+\mathcal{D}_ig(x)\mathcal{D}_jg(x)\mathcal{D}_{kl}g(x)\\
&+\mathcal{D}_jg(x)\mathcal{D}_{ik}g(x)\mathcal{D}_lg(x)+\mathcal{D}_ig(x)\mathcal{D}_{jk}g(x)\mathcal{D}_lg(x)+\mathcal{D}_kg(x)\mathcal{D}_{ij}g(x)\mathcal{D}_lg(x),\\
h_{44}(x)&=\mathcal{D}_ig(x)\mathcal{D}_jg(x)\mathcal{D}_kg(x)\mathcal{D}g_l(x).
\end{align*}

\subsubsection{Quadratic Forms}\label{quadratic-forms}

If \(g(x)=x'Ax\) for some symmetric \(A\), as it is in MDS, then \(\mathcal{D}_ig(x)=2\{Ax\}_i\) and \(\mathcal{D}_{ij}g(x)=2a_{ij}\). Also \(\mathcal{D}_{ijk}g(x)=0\) and \(\mathcal{D}_{ijkl}g(x)=0\). Note that we write \(\{Ax\}_i\) for element \(i\) of the vector \(Ax\). Also note that the results in this section apply for any symmetric matrix \(A\), even if it is not positive-semidefinite.

For the first partials
\[
\mathcal{D}_ih(x)=2\mathcal{D}f(x'Ax)\{Ax\}_i,
\]
for the second partials
\[
\mathcal{D}_{ij}h(x)=2\mathcal{D}f(x'Ax)a_{ij}+4\mathcal{D}^2f(x'Ax))\{Ax\}_i\{Ax\}_j,
\]
and for the third partials
\[
\mathcal{D}_{ijk}h(x)=4\mathcal{D}^2f(x'Ax)\{\{Ax\}_ia_{jk}+\{Ax\}_ja_{ik}+\{Ax\}_ka_{ij}\}+8\mathcal{D}^3f(x'Ax)\{Ax\}_i\{Ax\}_j\{Ax\}_k.
\]
For the fourth partials we again write
\[
\mathcal{D}_{ijkl}h(x)=\mathcal{D}f(g(x))h_{41}(x)+\mathcal{D}^2f(g(x))h_{42}(x)+\mathcal{D}^3f(g(x))h_{43}(x)+\mathcal{D}^4f(g(x))h_{44}(x),
\]
and now we have
\begin{align*}
h_{41}(x)&:=0,\\
h_{42}(x)&:=4\{a_{jl}a_{ik}+a_{il}a_{jk}\},\\
h_{43}(x)&:=8\{a_{il}\{Ax\}_j\{Ax\}_k+a_{jl}\{Ax\}_i\{Ax\}_k+a_{kl}\{Ax\}_i\{Ax\}_j+\\
&+a_{ik}\{Ax\}_j\{Ax\}_l+a_{jk}\{Ax\}_i\{Ax\}_l+a_{ij}\{Ax\}_k\{Ax\}_l\},\\
h_{44}(x)&:=16\{Ax\}_i\{Ax\}_j\{Ax\}_k\{Ax\}_l.
\end{align*}

\subsection{Derivatives of Stress}\label{derivatives-of-stress}

If we expand fStress we find, using notation used initially by De Leeuw (1977),
\[
\sigma(x)=C-\rho(x)+\eta(x),
\]
with
\[
\rho(x):=\mathop{\sum\sum}_{1\leq i<j\leq n}w_{ij}\delta_{ij}f(x'A_{ij}x),
\]
\[
\eta(x):=\frac12\mathop{\sum\sum}_{1\leq i<j\leq n}w_{ij}f^2(x'A_{ij}x),
\]
and with \(C\) a constant not depending on \(x\).

Clearly both \(\rho\) and \(\eta^2\) are weighted sums of fDistances, where the fDistances in \(\eta\) are the squares of the fDistances
in \(\rho\). Thus the partial derivatives are also weighted sums of the partial derivatives of the fDistances. Thus the previous formulas give us all we need.

\section{Implementation}\label{implementation}

As in many of our previous recent reports, most of the computing is done in a C dialect conforming with the .C() interface of R,
but also fairly easible integrated into C programs that do not depend on the presence of R.

At the moment we have implemented the following fDistances, which all are arbitrary powers of five base functions.
\begin{align}
h(x)&=\{\log(x'Ax)\}^p,\\
h(x)&=\{x'Ax\}^p,\\
h(x)&=\{\exp(x'Ax)\}^p,\\
h(x)&=\left\{\frac{x'Ax}{1+x'Ax}\right\}^p,\\
h(x)&=\{\log(1+x'Ax)\}^p.
\end{align}
There are helper functions in C which evaluate the base functions and all four of their partials at a given point. There is also
R glue for each of the helper functions, and the R function \texttt{checkD()} uses the symbolic differentiation capabilities of R to check the helper functions.

Partial derivatives of fDistances are evaluated using the multivariate Faà di Bruno results in the previous sections for the quadratic case with \(g(x)=x'Ax\). The derivatives of the powers of the five base functions are computed from the derivatives of the base functions, using the C function \texttt{fStressPower()}, which implements a univariate version of Faà di Bruno. Since our formulas cover arbitrary powers of the five base functions they apply to both the fDistances in \(\rho\) and their squares in \(\eta\).

The two basic C functions are
\texttt{faa\_di\_bruno()} and \texttt{fStressFaaDiBruno()}. The first computes the first four partials of \(f(x'Ax)\) for general \(A\), the second those
of \(f(x'Ax)\) for the \(A_{ij}\) used in MDS. In the second case we do not have to store \(A\) because there is a simple recipe to compute
its elements whenever they are needed. Our C implementation is far from optimal, because we store full multidimensional arrays of partials without using their super-symmetry. We also fill the arrays with full loops over all indices, without using the sparseness of
the \(A_{ij}\).

The R function \texttt{checkF()} checks the \texttt{faa\_di\_bruno()} derivatives using numeric differentiation (Gilbert and Varadhan (2019)), and the R function \texttt{faaR()} provides the glue for \texttt{faa\_di\_bruno}.

Finally there is the C function \texttt{fStressPartials()}, which makes the weighted sum of the partials of the fDistances. It has the R glue function \texttt{fStressR()}, and for the gradient and the Hessian it is checked against numerical derivatives by \texttt{checkS()}. In our examples the
first and second partials from \texttt{fStressR()} are the same as those from the \texttt{numDeriv} package. Since the higher derivatives of the
base functions and their powers given the same results as those from \texttt{deriv()} and \texttt{D()}, we can be reasonable sure that the higher derivatives are also correct. Checking this, and possibly improving the efficiency of the code, are topics for further research (and for a future version of this paper).

\section{Discussion}\label{discussion}

There is clearly some merit to the idea of fDistances, and to the general form of fStress. It covers many of the previous forms of least squares MDS. The first and second partials are not new. They can be used in gradient, Gauss-Newton, and Newton-Raphson minimization algorithms and to draw pseudo-confidence ellipsoids (De Leeuw (2017)). The third and fourth order partials allow for a better local approximation of fStress, which at least theoretically can lead to faster algorithms and more precise confidence regions.

In future research we will try to exploit convexity and concavity, and bounds on the derivatives, of fDistances and the implications for majorization algorithms to minimize fStress.

\section{Appendix: Code}\label{appendix-code}

\subsection{R Code}\label{r-code}

\subsubsection{fStress.R}\label{fstress.r}

\begin{Shaded}
\begin{Highlighting}[]
\FunctionTok{dyn.load}\NormalTok{(}\StringTok{"fStress.so"}\NormalTok{)}

\NormalTok{fLogR }\OtherTok{\textless{}{-}} \ControlFlowTok{function}\NormalTok{ (x) \{}
\NormalTok{  h }\OtherTok{\textless{}{-}}
    \FunctionTok{.C}\NormalTok{(}
      \StringTok{"fStressLog"}\NormalTok{,}
      \AttributeTok{x =} \FunctionTok{as.double}\NormalTok{ (x),}
      \AttributeTok{f0 =} \FunctionTok{as.double}\NormalTok{(}\DecValTok{0}\NormalTok{),}
      \AttributeTok{f1 =} \FunctionTok{as.double}\NormalTok{(}\DecValTok{0}\NormalTok{),}
      \AttributeTok{f2 =} \FunctionTok{as.double}\NormalTok{(}\DecValTok{0}\NormalTok{),}
      \AttributeTok{f3 =} \FunctionTok{as.double}\NormalTok{(}\DecValTok{0}\NormalTok{),}
      \AttributeTok{f4 =} \FunctionTok{as.double}\NormalTok{(}\DecValTok{0}\NormalTok{)}
\NormalTok{    )}
  \FunctionTok{return}\NormalTok{ (}\FunctionTok{list}\NormalTok{ (}
    \AttributeTok{x =}\NormalTok{ h}\SpecialCharTok{$}\NormalTok{x,}
    \AttributeTok{f0 =}\NormalTok{ h}\SpecialCharTok{$}\NormalTok{f0,}
    \AttributeTok{f1 =}\NormalTok{ h}\SpecialCharTok{$}\NormalTok{f1,}
    \AttributeTok{f2 =}\NormalTok{ h}\SpecialCharTok{$}\NormalTok{f2,}
    \AttributeTok{f3 =}\NormalTok{ h}\SpecialCharTok{$}\NormalTok{f3,}
    \AttributeTok{f4 =}\NormalTok{ h}\SpecialCharTok{$}\NormalTok{f4}
\NormalTok{  ))}
\NormalTok{\}}

\NormalTok{fIdeR }\OtherTok{\textless{}{-}} \ControlFlowTok{function}\NormalTok{ (x) \{}
\NormalTok{  h }\OtherTok{\textless{}{-}}
    \FunctionTok{.C}\NormalTok{(}
      \StringTok{"fStressIdentity"}\NormalTok{,}
      \AttributeTok{x =} \FunctionTok{as.double}\NormalTok{ (x),}
      \AttributeTok{f0 =} \FunctionTok{as.double}\NormalTok{(}\DecValTok{0}\NormalTok{),}
      \AttributeTok{f1 =} \FunctionTok{as.double}\NormalTok{(}\DecValTok{0}\NormalTok{),}
      \AttributeTok{f2 =} \FunctionTok{as.double}\NormalTok{(}\DecValTok{0}\NormalTok{),}
      \AttributeTok{f3 =} \FunctionTok{as.double}\NormalTok{(}\DecValTok{0}\NormalTok{),}
      \AttributeTok{f4 =} \FunctionTok{as.double}\NormalTok{(}\DecValTok{0}\NormalTok{)}
\NormalTok{    )}
  \FunctionTok{return}\NormalTok{ (}\FunctionTok{list}\NormalTok{ (}
    \AttributeTok{x =}\NormalTok{ h}\SpecialCharTok{$}\NormalTok{x,}
    \AttributeTok{f0 =}\NormalTok{ h}\SpecialCharTok{$}\NormalTok{f0,}
    \AttributeTok{f1 =}\NormalTok{ h}\SpecialCharTok{$}\NormalTok{f1,}
    \AttributeTok{f2 =}\NormalTok{ h}\SpecialCharTok{$}\NormalTok{f2,}
    \AttributeTok{f3 =}\NormalTok{ h}\SpecialCharTok{$}\NormalTok{f3,}
    \AttributeTok{f4 =}\NormalTok{ h}\SpecialCharTok{$}\NormalTok{f4}
\NormalTok{  ))}
\NormalTok{\}}

\NormalTok{fExpR }\OtherTok{\textless{}{-}} \ControlFlowTok{function}\NormalTok{ (x) \{}
\NormalTok{  h }\OtherTok{\textless{}{-}}
    \FunctionTok{.C}\NormalTok{(}
      \StringTok{"fStressExponent"}\NormalTok{,}
      \AttributeTok{x =} \FunctionTok{as.double}\NormalTok{ (x),}
      \AttributeTok{f0 =} \FunctionTok{as.double}\NormalTok{(}\DecValTok{0}\NormalTok{),}
      \AttributeTok{f1 =} \FunctionTok{as.double}\NormalTok{(}\DecValTok{0}\NormalTok{),}
      \AttributeTok{f2 =} \FunctionTok{as.double}\NormalTok{(}\DecValTok{0}\NormalTok{),}
      \AttributeTok{f3 =} \FunctionTok{as.double}\NormalTok{(}\DecValTok{0}\NormalTok{),}
      \AttributeTok{f4 =} \FunctionTok{as.double}\NormalTok{(}\DecValTok{0}\NormalTok{)}
\NormalTok{    )}
  \FunctionTok{return}\NormalTok{ (}\FunctionTok{list}\NormalTok{ (}
    \AttributeTok{x =}\NormalTok{ h}\SpecialCharTok{$}\NormalTok{x,}
    \AttributeTok{f0 =}\NormalTok{ h}\SpecialCharTok{$}\NormalTok{f0,}
    \AttributeTok{f1 =}\NormalTok{ h}\SpecialCharTok{$}\NormalTok{f1,}
    \AttributeTok{f2 =}\NormalTok{ h}\SpecialCharTok{$}\NormalTok{f2,}
    \AttributeTok{f3 =}\NormalTok{ h}\SpecialCharTok{$}\NormalTok{f3,}
    \AttributeTok{f4 =}\NormalTok{ h}\SpecialCharTok{$}\NormalTok{f4}
\NormalTok{  ))}
\NormalTok{\}}

\NormalTok{fBndR }\OtherTok{\textless{}{-}} \ControlFlowTok{function}\NormalTok{ (x) \{}
\NormalTok{  h }\OtherTok{\textless{}{-}}
    \FunctionTok{.C}\NormalTok{(}
      \StringTok{"fStressBounded"}\NormalTok{,}
      \AttributeTok{x =} \FunctionTok{as.double}\NormalTok{ (x),}
      \AttributeTok{f0 =} \FunctionTok{as.double}\NormalTok{(}\DecValTok{0}\NormalTok{),}
      \AttributeTok{f1 =} \FunctionTok{as.double}\NormalTok{(}\DecValTok{0}\NormalTok{),}
      \AttributeTok{f2 =} \FunctionTok{as.double}\NormalTok{(}\DecValTok{0}\NormalTok{),}
      \AttributeTok{f3 =} \FunctionTok{as.double}\NormalTok{(}\DecValTok{0}\NormalTok{),}
      \AttributeTok{f4 =} \FunctionTok{as.double}\NormalTok{(}\DecValTok{0}\NormalTok{)}
\NormalTok{    )}
  \FunctionTok{return}\NormalTok{ (}\FunctionTok{list}\NormalTok{ (}
    \AttributeTok{x =}\NormalTok{ h}\SpecialCharTok{$}\NormalTok{x,}
    \AttributeTok{f0 =}\NormalTok{ h}\SpecialCharTok{$}\NormalTok{f0,}
    \AttributeTok{f1 =}\NormalTok{ h}\SpecialCharTok{$}\NormalTok{f1,}
    \AttributeTok{f2 =}\NormalTok{ h}\SpecialCharTok{$}\NormalTok{f2,}
    \AttributeTok{f3 =}\NormalTok{ h}\SpecialCharTok{$}\NormalTok{f3,}
    \AttributeTok{f4 =}\NormalTok{ h}\SpecialCharTok{$}\NormalTok{f4}
\NormalTok{  ))}
\NormalTok{\}}

\NormalTok{fLgoR }\OtherTok{\textless{}{-}} \ControlFlowTok{function}\NormalTok{ (x) \{}
\NormalTok{  h }\OtherTok{\textless{}{-}}
    \FunctionTok{.C}\NormalTok{(}
      \StringTok{"fStressLogPlusOne"}\NormalTok{,}
      \AttributeTok{x =} \FunctionTok{as.double}\NormalTok{ (x),}
      \AttributeTok{f0 =} \FunctionTok{as.double}\NormalTok{(}\DecValTok{0}\NormalTok{),}
      \AttributeTok{f1 =} \FunctionTok{as.double}\NormalTok{(}\DecValTok{0}\NormalTok{),}
      \AttributeTok{f2 =} \FunctionTok{as.double}\NormalTok{(}\DecValTok{0}\NormalTok{),}
      \AttributeTok{f3 =} \FunctionTok{as.double}\NormalTok{(}\DecValTok{0}\NormalTok{),}
      \AttributeTok{f4 =} \FunctionTok{as.double}\NormalTok{(}\DecValTok{0}\NormalTok{)}
\NormalTok{    )}
  \FunctionTok{return}\NormalTok{ (}\FunctionTok{list}\NormalTok{ (}
    \AttributeTok{x =}\NormalTok{ h}\SpecialCharTok{$}\NormalTok{x,}
    \AttributeTok{f0 =}\NormalTok{ h}\SpecialCharTok{$}\NormalTok{f0,}
    \AttributeTok{f1 =}\NormalTok{ h}\SpecialCharTok{$}\NormalTok{f1,}
    \AttributeTok{f2 =}\NormalTok{ h}\SpecialCharTok{$}\NormalTok{f2,}
    \AttributeTok{f3 =}\NormalTok{ h}\SpecialCharTok{$}\NormalTok{f3,}
    \AttributeTok{f4 =}\NormalTok{ h}\SpecialCharTok{$}\NormalTok{f4}
\NormalTok{  ))}
\NormalTok{\}}

\NormalTok{fPowR }\OtherTok{\textless{}{-}} \ControlFlowTok{function}\NormalTok{ (x, fNumber, pPower) \{}
\NormalTok{  h }\OtherTok{\textless{}{-}}
    \FunctionTok{.C}\NormalTok{(}
      \StringTok{"fStressPower"}\NormalTok{,}
      \AttributeTok{x =} \FunctionTok{as.double}\NormalTok{(x),}
      \AttributeTok{fNumber =} \FunctionTok{as.integer}\NormalTok{ (fNumber),}
      \AttributeTok{ppower =} \FunctionTok{as.double}\NormalTok{ (pPower),}
      \AttributeTok{f0 =} \FunctionTok{as.double}\NormalTok{(}\DecValTok{0}\NormalTok{),}
      \AttributeTok{f1 =} \FunctionTok{as.double}\NormalTok{(}\DecValTok{0}\NormalTok{),}
      \AttributeTok{f2 =} \FunctionTok{as.double}\NormalTok{(}\DecValTok{0}\NormalTok{),}
      \AttributeTok{f3 =} \FunctionTok{as.double}\NormalTok{(}\DecValTok{0}\NormalTok{),}
      \AttributeTok{f4 =} \FunctionTok{as.double}\NormalTok{(}\DecValTok{0}\NormalTok{)}
\NormalTok{    )}
  \FunctionTok{return}\NormalTok{ (}\FunctionTok{list}\NormalTok{ (}
    \AttributeTok{x =}\NormalTok{ h}\SpecialCharTok{$}\NormalTok{x,}
    \AttributeTok{f0 =}\NormalTok{ h}\SpecialCharTok{$}\NormalTok{f0,}
    \AttributeTok{f1 =}\NormalTok{ h}\SpecialCharTok{$}\NormalTok{f1,}
    \AttributeTok{f2 =}\NormalTok{ h}\SpecialCharTok{$}\NormalTok{f2,}
    \AttributeTok{f3 =}\NormalTok{ h}\SpecialCharTok{$}\NormalTok{f3,}
    \AttributeTok{f4 =}\NormalTok{ h}\SpecialCharTok{$}\NormalTok{f4}
\NormalTok{  ))}
\NormalTok{\}}

\NormalTok{faaR }\OtherTok{\textless{}{-}} \ControlFlowTok{function}\NormalTok{(x, fNumber, pPower, a) \{}
\NormalTok{  n }\OtherTok{\textless{}{-}} \FunctionTok{length}\NormalTok{ (x)}
\NormalTok{  h }\OtherTok{\textless{}{-}}
    \FunctionTok{.C}\NormalTok{(}
      \StringTok{"faa\_di\_bruno"}\NormalTok{,}
      \AttributeTok{x =} \FunctionTok{as.double}\NormalTok{(x),}
      \AttributeTok{n =} \FunctionTok{as.integer}\NormalTok{(n),}
      \AttributeTok{fNumber =} \FunctionTok{as.integer}\NormalTok{(fNumber }\SpecialCharTok{{-}} \DecValTok{1}\NormalTok{),}
      \AttributeTok{pPower =} \FunctionTok{as.double}\NormalTok{(pPower),}
      \AttributeTok{a =} \FunctionTok{as.double}\NormalTok{(a),}
      \AttributeTok{ax =} \FunctionTok{as.double}\NormalTok{ (}\FunctionTok{rep}\NormalTok{(}\DecValTok{0}\NormalTok{, n)),}
      \AttributeTok{gx =} \FunctionTok{as.double}\NormalTok{(}\DecValTok{0}\NormalTok{),}
      \AttributeTok{h0 =} \FunctionTok{as.double}\NormalTok{ (}\DecValTok{0}\NormalTok{),}
      \AttributeTok{h1 =} \FunctionTok{as.double}\NormalTok{(}\FunctionTok{rep}\NormalTok{ (}\DecValTok{0}\NormalTok{, n)),}
      \AttributeTok{h2 =} \FunctionTok{as.double}\NormalTok{ (}\FunctionTok{rep}\NormalTok{(}\DecValTok{0}\NormalTok{, n }\SpecialCharTok{\^{}} \DecValTok{2}\NormalTok{)),}
      \AttributeTok{h3 =} \FunctionTok{as.double}\NormalTok{ (}\FunctionTok{rep}\NormalTok{ (}\DecValTok{0}\NormalTok{, n }\SpecialCharTok{\^{}} \DecValTok{3}\NormalTok{)),}
      \AttributeTok{h4 =} \FunctionTok{as.double}\NormalTok{ (}\FunctionTok{rep}\NormalTok{(}\DecValTok{0}\NormalTok{, n }\SpecialCharTok{\^{}} \DecValTok{4}\NormalTok{))}
\NormalTok{    )}
  \FunctionTok{return}\NormalTok{ (}\FunctionTok{list}\NormalTok{ (}
    \AttributeTok{gx =}\NormalTok{ h}\SpecialCharTok{$}\NormalTok{gx,}
    \AttributeTok{h0 =}\NormalTok{ h}\SpecialCharTok{$}\NormalTok{h0,}
    \AttributeTok{h1 =}\NormalTok{ h}\SpecialCharTok{$}\NormalTok{h1,}
    \AttributeTok{h2 =}\NormalTok{ h}\SpecialCharTok{$}\NormalTok{h2,}
    \AttributeTok{h3 =}\NormalTok{ h}\SpecialCharTok{$}\NormalTok{h3,}
    \AttributeTok{h4 =}\NormalTok{ h}\SpecialCharTok{$}\NormalTok{h4}
\NormalTok{  ))}
\NormalTok{\}}

\NormalTok{fStressR }\OtherTok{\textless{}{-}} \ControlFlowTok{function}\NormalTok{ (x, w, delta, p, fNumber, pPower) \{}
\NormalTok{  r }\OtherTok{\textless{}{-}} \FunctionTok{length}\NormalTok{ (x)}
\NormalTok{  m }\OtherTok{\textless{}{-}} \FunctionTok{length}\NormalTok{ (w)}
\NormalTok{  n }\OtherTok{\textless{}{-}} \FunctionTok{round}\NormalTok{ (r }\SpecialCharTok{/}\NormalTok{ p)}
\NormalTok{  hh }\OtherTok{\textless{}{-}}
    \FunctionTok{.C}\NormalTok{(}
      \StringTok{"fStressPartials"}\NormalTok{,}
      \AttributeTok{x =} \FunctionTok{as.double}\NormalTok{ (x),}
      \AttributeTok{w =} \FunctionTok{as.double}\NormalTok{ (w),}
      \AttributeTok{delta =} \FunctionTok{as.double}\NormalTok{ (delta),}
      \AttributeTok{n =} \FunctionTok{as.integer}\NormalTok{ (n),}
      \AttributeTok{p =} \FunctionTok{as.integer}\NormalTok{ (p),}
      \AttributeTok{fNumber =} \FunctionTok{as.integer}\NormalTok{ (fNumber }\SpecialCharTok{{-}} \DecValTok{1}\NormalTok{),}
      \AttributeTok{pPower =} \FunctionTok{as.double}\NormalTok{ (pPower),}
      \AttributeTok{stress =} \FunctionTok{as.double}\NormalTok{ (}\DecValTok{0}\NormalTok{),}
      \AttributeTok{qdist =} \FunctionTok{as.double}\NormalTok{ (}\FunctionTok{rep}\NormalTok{(}\DecValTok{0}\NormalTok{, m)),}
      \AttributeTok{fdist =} \FunctionTok{as.double}\NormalTok{ (}\FunctionTok{rep}\NormalTok{(}\DecValTok{0}\NormalTok{, m)),}
      \AttributeTok{h1 =} \FunctionTok{as.double}\NormalTok{ (}\FunctionTok{rep}\NormalTok{(}\DecValTok{0}\NormalTok{, r)),}
      \AttributeTok{h2 =} \FunctionTok{as.double}\NormalTok{ (}\FunctionTok{rep}\NormalTok{(}\DecValTok{0}\NormalTok{, r }\SpecialCharTok{\^{}} \DecValTok{2}\NormalTok{)),}
      \AttributeTok{h3 =} \FunctionTok{as.double}\NormalTok{ (}\FunctionTok{rep}\NormalTok{(}\DecValTok{0}\NormalTok{, r }\SpecialCharTok{\^{}} \DecValTok{3}\NormalTok{)),}
      \AttributeTok{h4 =} \FunctionTok{as.double}\NormalTok{ (}\FunctionTok{rep}\NormalTok{(}\DecValTok{0}\NormalTok{, r }\SpecialCharTok{\^{}} \DecValTok{4}\NormalTok{))}
\NormalTok{    )}
  \FunctionTok{return}\NormalTok{ (}
    \FunctionTok{list}\NormalTok{(}
      \AttributeTok{x =}\NormalTok{ x,}
      \AttributeTok{stress =}\NormalTok{ hh}\SpecialCharTok{$}\NormalTok{stress,}
      \AttributeTok{qdist =}\NormalTok{ hh}\SpecialCharTok{$}\NormalTok{qdist,}
      \AttributeTok{fdist =}\NormalTok{ hh}\SpecialCharTok{$}\NormalTok{fdist,}
      \AttributeTok{h1 =}\NormalTok{ hh}\SpecialCharTok{$}\NormalTok{h1,}
      \AttributeTok{h2 =}\NormalTok{ hh}\SpecialCharTok{$}\NormalTok{h2,}
      \AttributeTok{h3 =}\NormalTok{ hh}\SpecialCharTok{$}\NormalTok{h3,}
      \AttributeTok{h4 =}\NormalTok{ hh}\SpecialCharTok{$}\NormalTok{h4}
\NormalTok{    )}
\NormalTok{  )}
\NormalTok{\}}
\end{Highlighting}
\end{Shaded}

\subsubsection{check.R}\label{check.r}

\begin{Shaded}
\begin{Highlighting}[]
\FunctionTok{library}\NormalTok{(}\StringTok{"numDeriv"}\NormalTok{)}

\NormalTok{fList }\OtherTok{\textless{}{-}} \FunctionTok{list}\NormalTok{ (}\ControlFlowTok{function}\NormalTok{ (x)}
  \FunctionTok{log}\NormalTok{(x),}
  \ControlFlowTok{function}\NormalTok{ (x)}
\NormalTok{    x,}
  \ControlFlowTok{function}\NormalTok{ (x)}
    \FunctionTok{exp}\NormalTok{(x),}
  \ControlFlowTok{function}\NormalTok{ (x)}
\NormalTok{    x }\SpecialCharTok{/}\NormalTok{ (}\DecValTok{1} \SpecialCharTok{+}\NormalTok{ x),}
  \ControlFlowTok{function}\NormalTok{ (x)}
    \FunctionTok{log}\NormalTok{ (}\DecValTok{1} \SpecialCharTok{+}\NormalTok{ x))}

\NormalTok{checkD }\OtherTok{\textless{}{-}} \ControlFlowTok{function}\NormalTok{ (expr, order, value) \{}
\NormalTok{  DD }\OtherTok{\textless{}{-}} \ControlFlowTok{function}\NormalTok{(expr, name, }\AttributeTok{order =} \DecValTok{1}\NormalTok{) \{}
    \ControlFlowTok{if}\NormalTok{ (order }\SpecialCharTok{\textless{}} \DecValTok{1}\NormalTok{)}
      \FunctionTok{stop}\NormalTok{(}\StringTok{"\textquotesingle{}order\textquotesingle{} must be \textgreater{}= 1"}\NormalTok{)}
    \ControlFlowTok{if}\NormalTok{ (order }\SpecialCharTok{==} \DecValTok{1}\NormalTok{)}
      \FunctionTok{D}\NormalTok{(expr, name)}
    \ControlFlowTok{else}
      \FunctionTok{DD}\NormalTok{(}\FunctionTok{D}\NormalTok{(expr, name), name, order }\SpecialCharTok{{-}} \DecValTok{1}\NormalTok{)}
\NormalTok{  \}}
\NormalTok{  dd }\OtherTok{\textless{}{-}} \FunctionTok{DD}\NormalTok{(expr, }\StringTok{"x"}\NormalTok{, order)}
\NormalTok{  x }\OtherTok{\textless{}{-}}\NormalTok{ value}
  \FunctionTok{return}\NormalTok{ (}\FunctionTok{eval}\NormalTok{ (dd))}
\NormalTok{\}}

\NormalTok{checkF }\OtherTok{\textless{}{-}} \ControlFlowTok{function}\NormalTok{ (x, fNumber, pPower, a) \{}
\NormalTok{  f }\OtherTok{\textless{}{-}} \ControlFlowTok{function}\NormalTok{ (x, fNumber, pPower, a) \{}
\NormalTok{    g }\OtherTok{\textless{}{-}} \FunctionTok{sum}\NormalTok{ (a }\SpecialCharTok{*} \FunctionTok{outer}\NormalTok{ (x, x))}
    \FunctionTok{return}\NormalTok{ ((fList[[fNumber]](g)) }\SpecialCharTok{\^{}}\NormalTok{ pPower)}
\NormalTok{  \}}
\NormalTok{  h0 }\OtherTok{\textless{}{-}} \FunctionTok{f}\NormalTok{(x, fNumber, pPower, a)}
\NormalTok{  h1 }\OtherTok{\textless{}{-}} \FunctionTok{grad}\NormalTok{ (f,}
\NormalTok{              x,}
              \AttributeTok{fNumber =}\NormalTok{ fNumber,}
              \AttributeTok{pPower =}\NormalTok{ pPower,}
              \AttributeTok{a =}\NormalTok{ a)}
\NormalTok{  h2 }\OtherTok{\textless{}{-}} \FunctionTok{hessian}\NormalTok{ (f,}
\NormalTok{                 x,}
                 \AttributeTok{fNumber =}\NormalTok{ fNumber,}
                 \AttributeTok{pPower =}\NormalTok{ pPower,}
                 \AttributeTok{a =}\NormalTok{ a)}
  \FunctionTok{return}\NormalTok{ (}\FunctionTok{list}\NormalTok{ (}
    \AttributeTok{x =}\NormalTok{ x,}
    \AttributeTok{h0 =}\NormalTok{ h0,}
    \AttributeTok{h1 =}\NormalTok{ h1,}
    \AttributeTok{h2 =}\NormalTok{ h2}
\NormalTok{  ))}
\NormalTok{\}}

\NormalTok{checkS }\OtherTok{\textless{}{-}} \ControlFlowTok{function}\NormalTok{ (x, w, delta, p, fNumber, pPower) \{}
\NormalTok{  f }\OtherTok{\textless{}{-}} \ControlFlowTok{function}\NormalTok{ (x, w, delta, p, fNumber, pPower) \{}
\NormalTok{    r }\OtherTok{\textless{}{-}} \FunctionTok{length}\NormalTok{ (x)}
\NormalTok{    n }\OtherTok{\textless{}{-}} \FunctionTok{round}\NormalTok{ (r }\SpecialCharTok{/}\NormalTok{ p)}
\NormalTok{    d }\OtherTok{\textless{}{-}} \FunctionTok{as.vector}\NormalTok{(}\FunctionTok{dist}\NormalTok{(}\FunctionTok{matrix}\NormalTok{(x, n, p))) }\SpecialCharTok{\^{}} \DecValTok{2}
\NormalTok{    e }\OtherTok{\textless{}{-}}\NormalTok{ fList[[fNumber]](d) }\SpecialCharTok{\^{}}\NormalTok{ pPower}
    \FunctionTok{sum}\NormalTok{ (w }\SpecialCharTok{*}\NormalTok{ (delta }\SpecialCharTok{{-}}\NormalTok{ e) }\SpecialCharTok{\^{}} \DecValTok{2}\NormalTok{) }\SpecialCharTok{/} \DecValTok{2}
\NormalTok{  \}}
\NormalTok{  h0 }\OtherTok{\textless{}{-}}
    \FunctionTok{f}\NormalTok{(}
\NormalTok{      x,}
      \AttributeTok{w =}\NormalTok{ w,}
      \AttributeTok{delta =}\NormalTok{ delta,}
      \AttributeTok{p =}\NormalTok{ p,}
      \AttributeTok{fNumber =}\NormalTok{ fNumber,}
      \AttributeTok{pPower =}\NormalTok{ pPower}
\NormalTok{    )}
\NormalTok{  h1 }\OtherTok{\textless{}{-}}
    \FunctionTok{grad}\NormalTok{ (}
\NormalTok{      f,}
\NormalTok{      x,}
      \AttributeTok{w =}\NormalTok{ w,}
      \AttributeTok{delta =}\NormalTok{ delta,}
      \AttributeTok{p =}\NormalTok{ p,}
      \AttributeTok{fNumber =}\NormalTok{ fNumber,}
      \AttributeTok{pPower =}\NormalTok{ pPower}
\NormalTok{    )}
\NormalTok{  h2 }\OtherTok{\textless{}{-}}
    \FunctionTok{hessian}\NormalTok{ (}
\NormalTok{      f,}
\NormalTok{      x,}
      \AttributeTok{w =}\NormalTok{ w,}
      \AttributeTok{delta =}\NormalTok{ delta,}
      \AttributeTok{p =}\NormalTok{ p,}
      \AttributeTok{fNumber =}\NormalTok{ fNumber,}
      \AttributeTok{pPower =}\NormalTok{ pPower}
\NormalTok{    )}
  \FunctionTok{return}\NormalTok{ (}\FunctionTok{list}\NormalTok{ (}
    \AttributeTok{x =}\NormalTok{ x,}
    \AttributeTok{h0 =}\NormalTok{ h0,}
    \AttributeTok{h1 =}\NormalTok{ h1,}
    \AttributeTok{h2 =}\NormalTok{ h2}
\NormalTok{  ))}
\NormalTok{\}}
\end{Highlighting}
\end{Shaded}

\subsection{C Code}\label{c-code}

\subsubsection{fStress.h}\label{fstress.h}

\begin{Shaded}
\begin{Highlighting}[]
\PreprocessorTok{\#ifndef FSTRESS\_H}
\PreprocessorTok{\#define FSTRESS\_H}

\PreprocessorTok{\#include }\ImportTok{\textless{}math.h\textgreater{}}
\PreprocessorTok{\#include }\ImportTok{\textless{}stdbool.h\textgreater{}}
\PreprocessorTok{\#include }\ImportTok{\textless{}stdio.h\textgreater{}}
\PreprocessorTok{\#include }\ImportTok{\textless{}stdlib.h\textgreater{}}

\DataTypeTok{static} \KeywordTok{inline} \DataTypeTok{int}\NormalTok{ VINDEX}\OperatorTok{(}\DataTypeTok{const} \DataTypeTok{int}\OperatorTok{);}
\DataTypeTok{static} \KeywordTok{inline} \DataTypeTok{int}\NormalTok{ MINDEX}\OperatorTok{(}\DataTypeTok{const} \DataTypeTok{int}\OperatorTok{,} \DataTypeTok{const} \DataTypeTok{int}\OperatorTok{,} \DataTypeTok{const} \DataTypeTok{int}\OperatorTok{);}
\DataTypeTok{static} \KeywordTok{inline} \DataTypeTok{int}\NormalTok{ SINDEX}\OperatorTok{(}\DataTypeTok{const} \DataTypeTok{int}\OperatorTok{,} \DataTypeTok{const} \DataTypeTok{int}\OperatorTok{,} \DataTypeTok{const} \DataTypeTok{int}\OperatorTok{);}
\DataTypeTok{static} \KeywordTok{inline} \DataTypeTok{int}\NormalTok{ TINDEX}\OperatorTok{(}\DataTypeTok{const} \DataTypeTok{int}\OperatorTok{,} \DataTypeTok{const} \DataTypeTok{int}\OperatorTok{,} \DataTypeTok{const} \DataTypeTok{int}\OperatorTok{);}
\DataTypeTok{static} \KeywordTok{inline} \DataTypeTok{int}\NormalTok{ AINDEX}\OperatorTok{(}\DataTypeTok{const} \DataTypeTok{int}\OperatorTok{,} \DataTypeTok{const} \DataTypeTok{int}\OperatorTok{,} \DataTypeTok{const} \DataTypeTok{int}\OperatorTok{,} \DataTypeTok{const} \DataTypeTok{int}\OperatorTok{,} \DataTypeTok{const} \DataTypeTok{int}\OperatorTok{);}
\DataTypeTok{static} \KeywordTok{inline} \DataTypeTok{int}\NormalTok{ CINDEX}\OperatorTok{(}\DataTypeTok{const} \DataTypeTok{int}\OperatorTok{,} \DataTypeTok{const} \DataTypeTok{int}\OperatorTok{,} \DataTypeTok{const} \DataTypeTok{int}\OperatorTok{,} \DataTypeTok{const} \DataTypeTok{int}\OperatorTok{,} \DataTypeTok{const} \DataTypeTok{int}\OperatorTok{,} \DataTypeTok{const} \DataTypeTok{int}\OperatorTok{,} \DataTypeTok{const} \DataTypeTok{int}\OperatorTok{);}

\DataTypeTok{static} \KeywordTok{inline} \DataTypeTok{double}\NormalTok{ SQUARE}\OperatorTok{(}\DataTypeTok{const} \DataTypeTok{double}\OperatorTok{);}
\DataTypeTok{static} \KeywordTok{inline} \DataTypeTok{double}\NormalTok{ THIRD}\OperatorTok{(}\DataTypeTok{const} \DataTypeTok{double}\OperatorTok{);}
\DataTypeTok{static} \KeywordTok{inline} \DataTypeTok{double}\NormalTok{ FOURTH}\OperatorTok{(}\DataTypeTok{const} \DataTypeTok{double}\OperatorTok{);}
\DataTypeTok{static} \KeywordTok{inline} \DataTypeTok{double}\NormalTok{ FIFTH}\OperatorTok{(}\DataTypeTok{const} \DataTypeTok{double}\OperatorTok{);}

\DataTypeTok{static} \KeywordTok{inline} \DataTypeTok{double}\NormalTok{ MAX}\OperatorTok{(}\DataTypeTok{const} \DataTypeTok{double}\OperatorTok{,} \DataTypeTok{const} \DataTypeTok{double}\OperatorTok{);}
\DataTypeTok{static} \KeywordTok{inline} \DataTypeTok{double}\NormalTok{ MIN}\OperatorTok{(}\DataTypeTok{const} \DataTypeTok{double}\OperatorTok{,} \DataTypeTok{const} \DataTypeTok{double}\OperatorTok{);}
\DataTypeTok{static} \KeywordTok{inline} \DataTypeTok{int}\NormalTok{ IMIN}\OperatorTok{(}\DataTypeTok{const} \DataTypeTok{int}\OperatorTok{,} \DataTypeTok{const} \DataTypeTok{int}\OperatorTok{);}
\DataTypeTok{static} \KeywordTok{inline} \DataTypeTok{int}\NormalTok{ IMAX}\OperatorTok{(}\DataTypeTok{const} \DataTypeTok{int}\OperatorTok{,} \DataTypeTok{const} \DataTypeTok{int}\OperatorTok{);}
\DataTypeTok{static} \KeywordTok{inline} \DataTypeTok{int}\NormalTok{ IMOD}\OperatorTok{(}\DataTypeTok{const} \DataTypeTok{int}\OperatorTok{,} \DataTypeTok{const} \DataTypeTok{int}\OperatorTok{);}

\DataTypeTok{static} \KeywordTok{inline} \DataTypeTok{int}\NormalTok{ ASEEK}\OperatorTok{(}\DataTypeTok{const} \DataTypeTok{int}\OperatorTok{,} \DataTypeTok{const} \DataTypeTok{int}\OperatorTok{,} \DataTypeTok{const} \DataTypeTok{int}\OperatorTok{,} \DataTypeTok{const} \DataTypeTok{int}\OperatorTok{,} \DataTypeTok{const} \DataTypeTok{int}\OperatorTok{,} \DataTypeTok{const} \DataTypeTok{int}\OperatorTok{);}

\DataTypeTok{static} \KeywordTok{inline} \DataTypeTok{int}\NormalTok{ VINDEX}\OperatorTok{(}\DataTypeTok{const} \DataTypeTok{int}\NormalTok{ i}\OperatorTok{)} \OperatorTok{\{} \ControlFlowTok{return}\NormalTok{ i }\OperatorTok{{-}} \DecValTok{1}\OperatorTok{;} \OperatorTok{\}}

\DataTypeTok{static} \KeywordTok{inline} \DataTypeTok{int}\NormalTok{ MINDEX}\OperatorTok{(}\DataTypeTok{const} \DataTypeTok{int}\NormalTok{ i}\OperatorTok{,} \DataTypeTok{const} \DataTypeTok{int}\NormalTok{ j}\OperatorTok{,} \DataTypeTok{const} \DataTypeTok{int}\NormalTok{ n}\OperatorTok{)} \OperatorTok{\{}
    \ControlFlowTok{return} \OperatorTok{(}\NormalTok{i }\OperatorTok{{-}} \DecValTok{1}\OperatorTok{)} \OperatorTok{+} \OperatorTok{(}\NormalTok{j }\OperatorTok{{-}} \DecValTok{1}\OperatorTok{)} \OperatorTok{*}\NormalTok{ n}\OperatorTok{;}
\OperatorTok{\}}

\DataTypeTok{static} \KeywordTok{inline} \DataTypeTok{int}\NormalTok{ AINDEX}\OperatorTok{(}\DataTypeTok{const} \DataTypeTok{int}\NormalTok{ i}\OperatorTok{,} \DataTypeTok{const} \DataTypeTok{int}\NormalTok{ j}\OperatorTok{,} \DataTypeTok{const} \DataTypeTok{int}\NormalTok{ k}\OperatorTok{,} \DataTypeTok{const} \DataTypeTok{int}\NormalTok{ n}\OperatorTok{,}
                         \DataTypeTok{const} \DataTypeTok{int}\NormalTok{ m}\OperatorTok{)} \OperatorTok{\{}
    \ControlFlowTok{return} \OperatorTok{(}\NormalTok{i }\OperatorTok{{-}} \DecValTok{1}\OperatorTok{)} \OperatorTok{+} \OperatorTok{(}\NormalTok{j }\OperatorTok{{-}} \DecValTok{1}\OperatorTok{)} \OperatorTok{*}\NormalTok{ n }\OperatorTok{+} \OperatorTok{(}\NormalTok{k }\OperatorTok{{-}} \DecValTok{1}\OperatorTok{)} \OperatorTok{*}\NormalTok{ n }\OperatorTok{*}\NormalTok{ m}\OperatorTok{;}
\OperatorTok{\}}

\DataTypeTok{static} \KeywordTok{inline} \DataTypeTok{int}\NormalTok{ CINDEX}\OperatorTok{(}\DataTypeTok{const} \DataTypeTok{int}\NormalTok{ i}\OperatorTok{,} \DataTypeTok{const} \DataTypeTok{int}\NormalTok{ j}\OperatorTok{,} \DataTypeTok{const} \DataTypeTok{int}\NormalTok{ k}\OperatorTok{,} \DataTypeTok{const} \DataTypeTok{int}\NormalTok{ l}\OperatorTok{,} \DataTypeTok{const} \DataTypeTok{int}\NormalTok{ n}\OperatorTok{,}
                         \DataTypeTok{const} \DataTypeTok{int}\NormalTok{ m}\OperatorTok{,} \DataTypeTok{const} \DataTypeTok{int}\NormalTok{ r}\OperatorTok{)} \OperatorTok{\{}
    \ControlFlowTok{return} \OperatorTok{(}\NormalTok{i }\OperatorTok{{-}} \DecValTok{1}\OperatorTok{)} \OperatorTok{+} \OperatorTok{(}\NormalTok{j }\OperatorTok{{-}} \DecValTok{1}\OperatorTok{)} \OperatorTok{*}\NormalTok{ n }\OperatorTok{+} \OperatorTok{(}\NormalTok{k }\OperatorTok{{-}} \DecValTok{1}\OperatorTok{)} \OperatorTok{*}\NormalTok{ n }\OperatorTok{*}\NormalTok{ m }\OperatorTok{+} \OperatorTok{(}\NormalTok{l }\OperatorTok{{-}} \DecValTok{1}\OperatorTok{)} \OperatorTok{*}\NormalTok{ n }\OperatorTok{*}\NormalTok{ m }\OperatorTok{*}\NormalTok{ r}\OperatorTok{;}
\OperatorTok{\}}

\DataTypeTok{static} \KeywordTok{inline} \DataTypeTok{int}\NormalTok{ SINDEX}\OperatorTok{(}\DataTypeTok{const} \DataTypeTok{int}\NormalTok{ i}\OperatorTok{,} \DataTypeTok{const} \DataTypeTok{int}\NormalTok{ j}\OperatorTok{,} \DataTypeTok{const} \DataTypeTok{int}\NormalTok{ n}\OperatorTok{)} \OperatorTok{\{}
    \ControlFlowTok{return} \OperatorTok{((}\NormalTok{j }\OperatorTok{{-}} \DecValTok{1}\OperatorTok{)} \OperatorTok{*}\NormalTok{ n}\OperatorTok{)} \OperatorTok{{-}} \OperatorTok{(}\NormalTok{j }\OperatorTok{*} \OperatorTok{(}\NormalTok{j }\OperatorTok{{-}} \DecValTok{1}\OperatorTok{)} \OperatorTok{/} \DecValTok{2}\OperatorTok{)} \OperatorTok{+} \OperatorTok{(}\NormalTok{i }\OperatorTok{{-}}\NormalTok{ j}\OperatorTok{)} \OperatorTok{{-}} \DecValTok{1}\OperatorTok{;}
\OperatorTok{\}}

\DataTypeTok{static} \KeywordTok{inline} \DataTypeTok{int}\NormalTok{ TINDEX}\OperatorTok{(}\DataTypeTok{const} \DataTypeTok{int}\NormalTok{ i}\OperatorTok{,} \DataTypeTok{const} \DataTypeTok{int}\NormalTok{ j}\OperatorTok{,} \DataTypeTok{const} \DataTypeTok{int}\NormalTok{ n}\OperatorTok{)} \OperatorTok{\{}
    \ControlFlowTok{return} \OperatorTok{((}\NormalTok{j }\OperatorTok{{-}} \DecValTok{1}\OperatorTok{)} \OperatorTok{*}\NormalTok{ n}\OperatorTok{)} \OperatorTok{{-}} \OperatorTok{((}\NormalTok{j }\OperatorTok{{-}} \DecValTok{1}\OperatorTok{)} \OperatorTok{*} \OperatorTok{(}\NormalTok{j }\OperatorTok{{-}} \DecValTok{2}\OperatorTok{)} \OperatorTok{/} \DecValTok{2}\OperatorTok{)} \OperatorTok{+} \OperatorTok{(}\NormalTok{i }\OperatorTok{{-}} \OperatorTok{(}\NormalTok{j }\OperatorTok{{-}} \DecValTok{1}\OperatorTok{))} \OperatorTok{{-}} \DecValTok{1}\OperatorTok{;}
\OperatorTok{\}}

\DataTypeTok{static} \KeywordTok{inline} \DataTypeTok{double}\NormalTok{ SQUARE}\OperatorTok{(}\DataTypeTok{const} \DataTypeTok{double}\NormalTok{ x}\OperatorTok{)} \OperatorTok{\{} \ControlFlowTok{return}\NormalTok{ x }\OperatorTok{*}\NormalTok{ x}\OperatorTok{;} \OperatorTok{\}}
\DataTypeTok{static} \KeywordTok{inline} \DataTypeTok{double}\NormalTok{ THIRD}\OperatorTok{(}\DataTypeTok{const} \DataTypeTok{double}\NormalTok{ x}\OperatorTok{)} \OperatorTok{\{} \ControlFlowTok{return}\NormalTok{ x }\OperatorTok{*}\NormalTok{ x }\OperatorTok{*}\NormalTok{ x}\OperatorTok{;} \OperatorTok{\}}
\DataTypeTok{static} \KeywordTok{inline} \DataTypeTok{double}\NormalTok{ FOURTH}\OperatorTok{(}\DataTypeTok{const} \DataTypeTok{double}\NormalTok{ x}\OperatorTok{)} \OperatorTok{\{} \ControlFlowTok{return}\NormalTok{ x }\OperatorTok{*}\NormalTok{ x }\OperatorTok{*}\NormalTok{ x }\OperatorTok{*}\NormalTok{ x}\OperatorTok{;} \OperatorTok{\}}
\DataTypeTok{static} \KeywordTok{inline} \DataTypeTok{double}\NormalTok{ FIFTH}\OperatorTok{(}\DataTypeTok{const} \DataTypeTok{double}\NormalTok{ x}\OperatorTok{)} \OperatorTok{\{} \ControlFlowTok{return}\NormalTok{ x }\OperatorTok{*}\NormalTok{ x }\OperatorTok{*}\NormalTok{ x }\OperatorTok{*}\NormalTok{ x }\OperatorTok{*}\NormalTok{ x}\OperatorTok{;} \OperatorTok{\}}

\DataTypeTok{static} \KeywordTok{inline} \DataTypeTok{double}\NormalTok{ MAX}\OperatorTok{(}\DataTypeTok{const} \DataTypeTok{double}\NormalTok{ x}\OperatorTok{,} \DataTypeTok{const} \DataTypeTok{double}\NormalTok{ y}\OperatorTok{)} \OperatorTok{\{}
    \ControlFlowTok{return} \OperatorTok{(}\NormalTok{x }\OperatorTok{\textgreater{}}\NormalTok{ y}\OperatorTok{)} \OperatorTok{?}\NormalTok{ x }\OperatorTok{:}\NormalTok{ y}\OperatorTok{;}
\OperatorTok{\}}

\DataTypeTok{static} \KeywordTok{inline} \DataTypeTok{double}\NormalTok{ MIN}\OperatorTok{(}\DataTypeTok{const} \DataTypeTok{double}\NormalTok{ x}\OperatorTok{,} \DataTypeTok{const} \DataTypeTok{double}\NormalTok{ y}\OperatorTok{)} \OperatorTok{\{}
    \ControlFlowTok{return} \OperatorTok{(}\NormalTok{x }\OperatorTok{\textless{}}\NormalTok{ y}\OperatorTok{)} \OperatorTok{?}\NormalTok{ x }\OperatorTok{:}\NormalTok{ y}\OperatorTok{;}
\OperatorTok{\}}

\DataTypeTok{static} \KeywordTok{inline} \DataTypeTok{int}\NormalTok{ IMAX}\OperatorTok{(}\DataTypeTok{const} \DataTypeTok{int}\NormalTok{ x}\OperatorTok{,} \DataTypeTok{const} \DataTypeTok{int}\NormalTok{ y}\OperatorTok{)} \OperatorTok{\{} \ControlFlowTok{return} \OperatorTok{(}\NormalTok{x }\OperatorTok{\textgreater{}}\NormalTok{ y}\OperatorTok{)} \OperatorTok{?}\NormalTok{ x }\OperatorTok{:}\NormalTok{ y}\OperatorTok{;} \OperatorTok{\}}

\DataTypeTok{static} \KeywordTok{inline} \DataTypeTok{int}\NormalTok{ IMIN}\OperatorTok{(}\DataTypeTok{const} \DataTypeTok{int}\NormalTok{ x}\OperatorTok{,} \DataTypeTok{const} \DataTypeTok{int}\NormalTok{ y}\OperatorTok{)} \OperatorTok{\{} \ControlFlowTok{return} \OperatorTok{(}\NormalTok{x }\OperatorTok{\textless{}}\NormalTok{ y}\OperatorTok{)} \OperatorTok{?}\NormalTok{ x }\OperatorTok{:}\NormalTok{ y}\OperatorTok{;} \OperatorTok{\}}

\DataTypeTok{static} \KeywordTok{inline} \DataTypeTok{int}\NormalTok{ IMOD}\OperatorTok{(}\DataTypeTok{const} \DataTypeTok{int}\NormalTok{ x}\OperatorTok{,} \DataTypeTok{const} \DataTypeTok{int}\NormalTok{ y}\OperatorTok{)} \OperatorTok{\{}
    \ControlFlowTok{return} \OperatorTok{(((}\NormalTok{x }\OperatorTok{\%}\NormalTok{ y}\OperatorTok{)} \OperatorTok{==} \DecValTok{0}\OperatorTok{)} \OperatorTok{?}\NormalTok{ y }\OperatorTok{:} \OperatorTok{(}\NormalTok{x }\OperatorTok{\%}\NormalTok{ y}\OperatorTok{));}
\OperatorTok{\}}

\DataTypeTok{static} \KeywordTok{inline} \DataTypeTok{int}\NormalTok{ ASEEK}\OperatorTok{(}\DataTypeTok{const} \DataTypeTok{int}\NormalTok{ n}\OperatorTok{,} \DataTypeTok{const} \DataTypeTok{int}\NormalTok{ p}\OperatorTok{,} \DataTypeTok{const} \DataTypeTok{int}\NormalTok{ u}\OperatorTok{,} \DataTypeTok{const} \DataTypeTok{int}\NormalTok{ v}\OperatorTok{,} \DataTypeTok{const} \DataTypeTok{int}\NormalTok{ i}\OperatorTok{,} \DataTypeTok{const} \DataTypeTok{int}\NormalTok{ j}\OperatorTok{)} \OperatorTok{\{}
    \DataTypeTok{int}\NormalTok{ value }\OperatorTok{=} \DecValTok{0}\OperatorTok{;}
    \ControlFlowTok{if} \OperatorTok{(}\NormalTok{abs }\OperatorTok{(}\NormalTok{i }\OperatorTok{{-}}\NormalTok{ j}\OperatorTok{)} \OperatorTok{\textless{}}\NormalTok{ n}\OperatorTok{)} \OperatorTok{\{}
        \ControlFlowTok{for} \OperatorTok{(}\DataTypeTok{int}\NormalTok{ s }\OperatorTok{=} \DecValTok{1}\OperatorTok{;}\NormalTok{ s }\OperatorTok{\textless{}=}\NormalTok{ p}\OperatorTok{;}\NormalTok{ s}\OperatorTok{++)} \OperatorTok{\{}
            \DataTypeTok{int}\NormalTok{ k }\OperatorTok{=} \OperatorTok{(}\NormalTok{s }\OperatorTok{{-}} \DecValTok{1}\OperatorTok{)} \OperatorTok{*}\NormalTok{ n}\OperatorTok{;}
            \ControlFlowTok{if} \OperatorTok{((}\NormalTok{i }\OperatorTok{==} \OperatorTok{(}\NormalTok{u }\OperatorTok{+}\NormalTok{ k}\OperatorTok{))} \OperatorTok{\&\&} \OperatorTok{(}\NormalTok{j }\OperatorTok{==} \OperatorTok{(}\NormalTok{u }\OperatorTok{+}\NormalTok{ k}\OperatorTok{)))} \OperatorTok{\{}
\NormalTok{                value }\OperatorTok{=} \DecValTok{1}\OperatorTok{;}
            \OperatorTok{\}}
            \ControlFlowTok{if} \OperatorTok{((}\NormalTok{i }\OperatorTok{==} \OperatorTok{(}\NormalTok{u }\OperatorTok{+}\NormalTok{ k}\OperatorTok{))} \OperatorTok{\&\&} \OperatorTok{(}\NormalTok{j }\OperatorTok{==} \OperatorTok{(}\NormalTok{v }\OperatorTok{+}\NormalTok{ k}\OperatorTok{)))} \OperatorTok{\{}
\NormalTok{                value }\OperatorTok{=} \OperatorTok{{-}}\DecValTok{1}\OperatorTok{;}
            \OperatorTok{\}}
            \ControlFlowTok{if} \OperatorTok{((}\NormalTok{i }\OperatorTok{==} \OperatorTok{(}\NormalTok{v }\OperatorTok{+}\NormalTok{ k}\OperatorTok{))} \OperatorTok{\&\&} \OperatorTok{(}\NormalTok{j }\OperatorTok{==} \OperatorTok{(}\NormalTok{u }\OperatorTok{+}\NormalTok{ k}\OperatorTok{)))} \OperatorTok{\{}
\NormalTok{                value }\OperatorTok{=} \OperatorTok{{-}}\DecValTok{1}\OperatorTok{;}
            \OperatorTok{\}}
            \ControlFlowTok{if} \OperatorTok{((}\NormalTok{i }\OperatorTok{==} \OperatorTok{(}\NormalTok{v }\OperatorTok{+}\NormalTok{ k}\OperatorTok{))} \OperatorTok{\&\&} \OperatorTok{(}\NormalTok{j }\OperatorTok{==} \OperatorTok{(}\NormalTok{v }\OperatorTok{+}\NormalTok{ k}\OperatorTok{)))} \OperatorTok{\{}
\NormalTok{                value }\OperatorTok{=} \DecValTok{1}\OperatorTok{;}
            \OperatorTok{\}}
        \OperatorTok{\}}
    \OperatorTok{\}}
    \ControlFlowTok{return} \OperatorTok{(}\NormalTok{value}\OperatorTok{);}
\OperatorTok{\}}


\PreprocessorTok{\#endif }\CommentTok{/* FSTRESS\_H */}
\end{Highlighting}
\end{Shaded}

\subsubsection{fStress.c}\label{fstress.c}

\begin{Shaded}
\begin{Highlighting}[]
\PreprocessorTok{\#include }\ImportTok{"fStress.h"}

\DataTypeTok{void}\NormalTok{ fStressLog}\OperatorTok{(}\DataTypeTok{const} \DataTypeTok{double} \OperatorTok{*}\NormalTok{x}\OperatorTok{,} \DataTypeTok{double} \OperatorTok{*}\NormalTok{f0}\OperatorTok{,} \DataTypeTok{double} \OperatorTok{*}\NormalTok{f1}\OperatorTok{,} \DataTypeTok{double} \OperatorTok{*}\NormalTok{f2}\OperatorTok{,} \DataTypeTok{double} \OperatorTok{*}\NormalTok{f3}\OperatorTok{,}
                \DataTypeTok{double} \OperatorTok{*}\NormalTok{f4}\OperatorTok{)} \OperatorTok{\{}
    \DataTypeTok{double}\NormalTok{ xx }\OperatorTok{=} \OperatorTok{*}\NormalTok{x}\OperatorTok{;}
    \OperatorTok{*}\NormalTok{f0 }\OperatorTok{=}\NormalTok{ log}\OperatorTok{(}\NormalTok{xx}\OperatorTok{);}
    \OperatorTok{*}\NormalTok{f1 }\OperatorTok{=} \DecValTok{1} \OperatorTok{/}\NormalTok{ xx}\OperatorTok{;}
    \OperatorTok{*}\NormalTok{f2 }\OperatorTok{=} \OperatorTok{{-}}\DecValTok{1} \OperatorTok{/}\NormalTok{ SQUARE}\OperatorTok{(}\NormalTok{xx}\OperatorTok{);}
    \OperatorTok{*}\NormalTok{f3 }\OperatorTok{=} \DecValTok{2} \OperatorTok{/}\NormalTok{ THIRD}\OperatorTok{(}\NormalTok{xx}\OperatorTok{);}
    \OperatorTok{*}\NormalTok{f4 }\OperatorTok{=} \OperatorTok{{-}}\DecValTok{6} \OperatorTok{/}\NormalTok{ FOURTH}\OperatorTok{(}\NormalTok{xx}\OperatorTok{);}
    \ControlFlowTok{return}\OperatorTok{;}
\OperatorTok{\}}

\DataTypeTok{void}\NormalTok{ fStressIdentity}\OperatorTok{(}\DataTypeTok{const} \DataTypeTok{double} \OperatorTok{*}\NormalTok{x}\OperatorTok{,} \DataTypeTok{double} \OperatorTok{*}\NormalTok{f0}\OperatorTok{,} \DataTypeTok{double} \OperatorTok{*}\NormalTok{f1}\OperatorTok{,} \DataTypeTok{double} \OperatorTok{*}\NormalTok{f2}\OperatorTok{,}
                     \DataTypeTok{double} \OperatorTok{*}\NormalTok{f3}\OperatorTok{,} \DataTypeTok{double} \OperatorTok{*}\NormalTok{f4}\OperatorTok{)} \OperatorTok{\{}
    \DataTypeTok{double}\NormalTok{ xx }\OperatorTok{=} \OperatorTok{*}\NormalTok{x}\OperatorTok{;}
    \OperatorTok{*}\NormalTok{f0 }\OperatorTok{=}\NormalTok{ xx}\OperatorTok{;}
    \OperatorTok{*}\NormalTok{f1 }\OperatorTok{=} \FloatTok{1.0}\OperatorTok{;}
    \OperatorTok{*}\NormalTok{f2 }\OperatorTok{=} \FloatTok{0.0}\OperatorTok{;}
    \OperatorTok{*}\NormalTok{f3 }\OperatorTok{=} \FloatTok{0.0}\OperatorTok{;}
    \OperatorTok{*}\NormalTok{f4 }\OperatorTok{=} \FloatTok{0.0}\OperatorTok{;}
    \ControlFlowTok{return}\OperatorTok{;}
\OperatorTok{\}}

\DataTypeTok{void}\NormalTok{ fStressExponent}\OperatorTok{(}\DataTypeTok{const} \DataTypeTok{double} \OperatorTok{*}\NormalTok{x}\OperatorTok{,} \DataTypeTok{double} \OperatorTok{*}\NormalTok{f0}\OperatorTok{,} \DataTypeTok{double} \OperatorTok{*}\NormalTok{f1}\OperatorTok{,} \DataTypeTok{double} \OperatorTok{*}\NormalTok{f2}\OperatorTok{,}
                     \DataTypeTok{double} \OperatorTok{*}\NormalTok{f3}\OperatorTok{,} \DataTypeTok{double} \OperatorTok{*}\NormalTok{f4}\OperatorTok{)} \OperatorTok{\{}
    \DataTypeTok{double}\NormalTok{ xx }\OperatorTok{=} \OperatorTok{*}\NormalTok{x}\OperatorTok{;}
    \OperatorTok{*}\NormalTok{f0 }\OperatorTok{=}\NormalTok{ exp}\OperatorTok{(}\NormalTok{xx}\OperatorTok{);}
    \OperatorTok{*}\NormalTok{f1 }\OperatorTok{=}\NormalTok{ exp}\OperatorTok{(}\NormalTok{xx}\OperatorTok{);}
    \OperatorTok{*}\NormalTok{f2 }\OperatorTok{=}\NormalTok{ exp}\OperatorTok{(}\NormalTok{xx}\OperatorTok{);}
    \OperatorTok{*}\NormalTok{f3 }\OperatorTok{=}\NormalTok{ exp}\OperatorTok{(}\NormalTok{xx}\OperatorTok{);}
    \OperatorTok{*}\NormalTok{f4 }\OperatorTok{=}\NormalTok{ exp}\OperatorTok{(}\NormalTok{xx}\OperatorTok{);}
    \ControlFlowTok{return}\OperatorTok{;}
\OperatorTok{\}}

\DataTypeTok{void}\NormalTok{ fStressBounded}\OperatorTok{(}\DataTypeTok{const} \DataTypeTok{double} \OperatorTok{*}\NormalTok{x}\OperatorTok{,} \DataTypeTok{double} \OperatorTok{*}\NormalTok{f0}\OperatorTok{,} \DataTypeTok{double} \OperatorTok{*}\NormalTok{f1}\OperatorTok{,} \DataTypeTok{double} \OperatorTok{*}\NormalTok{f2}\OperatorTok{,}
                    \DataTypeTok{double} \OperatorTok{*}\NormalTok{f3}\OperatorTok{,} \DataTypeTok{double} \OperatorTok{*}\NormalTok{f4}\OperatorTok{)} \OperatorTok{\{}
    \DataTypeTok{double}\NormalTok{ xx }\OperatorTok{=} \OperatorTok{*}\NormalTok{x}\OperatorTok{,}\NormalTok{ xz }\OperatorTok{=} \FloatTok{1.0} \OperatorTok{+}\NormalTok{ xx}\OperatorTok{;}
    \OperatorTok{*}\NormalTok{f0 }\OperatorTok{=}\NormalTok{ xx }\OperatorTok{/}\NormalTok{ xz}\OperatorTok{;}
    \OperatorTok{*}\NormalTok{f1 }\OperatorTok{=} \FloatTok{1.0} \OperatorTok{/}\NormalTok{ SQUARE}\OperatorTok{(}\NormalTok{xz}\OperatorTok{);}
    \OperatorTok{*}\NormalTok{f2 }\OperatorTok{=} \OperatorTok{{-}}\FloatTok{2.0} \OperatorTok{/}\NormalTok{ THIRD}\OperatorTok{(}\NormalTok{xz}\OperatorTok{);}
    \OperatorTok{*}\NormalTok{f3 }\OperatorTok{=} \FloatTok{6.0} \OperatorTok{/}\NormalTok{ FOURTH}\OperatorTok{(}\NormalTok{xz}\OperatorTok{);}
    \OperatorTok{*}\NormalTok{f4 }\OperatorTok{=} \OperatorTok{{-}}\FloatTok{24.0} \OperatorTok{/}\NormalTok{ FIFTH}\OperatorTok{(}\NormalTok{xz}\OperatorTok{);}
    \ControlFlowTok{return}\OperatorTok{;}
\OperatorTok{\}}

\DataTypeTok{void}\NormalTok{ fStressLogPlusOne}\OperatorTok{(}\DataTypeTok{const} \DataTypeTok{double} \OperatorTok{*}\NormalTok{x}\OperatorTok{,} \DataTypeTok{double} \OperatorTok{*}\NormalTok{f0}\OperatorTok{,} \DataTypeTok{double} \OperatorTok{*}\NormalTok{f1}\OperatorTok{,} \DataTypeTok{double} \OperatorTok{*}\NormalTok{f2}\OperatorTok{,}
                       \DataTypeTok{double} \OperatorTok{*}\NormalTok{f3}\OperatorTok{,} \DataTypeTok{double} \OperatorTok{*}\NormalTok{f4}\OperatorTok{)} \OperatorTok{\{}
    \DataTypeTok{double}\NormalTok{ xx }\OperatorTok{=} \OperatorTok{*}\NormalTok{x}\OperatorTok{,}\NormalTok{ xp }\OperatorTok{=} \FloatTok{1.0} \OperatorTok{+}\NormalTok{ xx}\OperatorTok{,}\NormalTok{ xz }\OperatorTok{=} \DecValTok{1} \OperatorTok{/}\NormalTok{ xp}\OperatorTok{;}
    \OperatorTok{*}\NormalTok{f0 }\OperatorTok{=}\NormalTok{ log}\OperatorTok{(}\NormalTok{xp}\OperatorTok{);}
    \OperatorTok{*}\NormalTok{f1 }\OperatorTok{=}\NormalTok{ xz}\OperatorTok{;}
    \OperatorTok{*}\NormalTok{f2 }\OperatorTok{=} \OperatorTok{{-}}\NormalTok{SQUARE}\OperatorTok{(}\NormalTok{xz}\OperatorTok{);}
    \OperatorTok{*}\NormalTok{f3 }\OperatorTok{=} \FloatTok{2.0} \OperatorTok{*}\NormalTok{ THIRD}\OperatorTok{(}\NormalTok{xz}\OperatorTok{);}
    \OperatorTok{*}\NormalTok{f4 }\OperatorTok{=} \OperatorTok{{-}}\FloatTok{6.0} \OperatorTok{*}\NormalTok{ FOURTH}\OperatorTok{(}\NormalTok{xz}\OperatorTok{);}
\OperatorTok{\}}

\DataTypeTok{void} \OperatorTok{(*}\NormalTok{fStressTable}\OperatorTok{[}\DecValTok{5}\OperatorTok{])(}\DataTypeTok{const} \DataTypeTok{double} \OperatorTok{*,} \DataTypeTok{double} \OperatorTok{*,} \DataTypeTok{double} \OperatorTok{*,} \DataTypeTok{double} \OperatorTok{*,} \DataTypeTok{double} \OperatorTok{*,}
                        \DataTypeTok{double} \OperatorTok{*)} \OperatorTok{=} \OperatorTok{\{}\NormalTok{fStressLog}\OperatorTok{,}\NormalTok{ fStressIdentity}\OperatorTok{,}
\NormalTok{                                     fStressExponent}\OperatorTok{,}\NormalTok{ fStressBounded}\OperatorTok{,}
\NormalTok{                                     fStressLogPlusOne}\OperatorTok{\};}

\DataTypeTok{void}\NormalTok{ fStressPower}\OperatorTok{(}\DataTypeTok{const} \DataTypeTok{double} \OperatorTok{*}\NormalTok{x}\OperatorTok{,} \DataTypeTok{const} \DataTypeTok{int} \OperatorTok{*}\NormalTok{fNumber}\OperatorTok{,} \DataTypeTok{const} \DataTypeTok{double} \OperatorTok{*}\NormalTok{ppar}\OperatorTok{,}
                  \DataTypeTok{double} \OperatorTok{*}\NormalTok{g0}\OperatorTok{,} \DataTypeTok{double} \OperatorTok{*}\NormalTok{g1}\OperatorTok{,} \DataTypeTok{double} \OperatorTok{*}\NormalTok{g2}\OperatorTok{,} \DataTypeTok{double} \OperatorTok{*}\NormalTok{g3}\OperatorTok{,} \DataTypeTok{double} \OperatorTok{*}\NormalTok{g4}\OperatorTok{)} \OperatorTok{\{}
    \DataTypeTok{double}\NormalTok{ f0}\OperatorTok{,}\NormalTok{ f1}\OperatorTok{,}\NormalTok{ f2}\OperatorTok{,}\NormalTok{ f3}\OperatorTok{,}\NormalTok{ f4}\OperatorTok{,}\NormalTok{ pp }\OperatorTok{=} \OperatorTok{*}\NormalTok{ppar}\OperatorTok{;}
    \OperatorTok{(}\DataTypeTok{void}\OperatorTok{)}\NormalTok{fStressTable}\OperatorTok{[*}\NormalTok{fNumber}\OperatorTok{](}\NormalTok{x}\OperatorTok{,} \OperatorTok{\&}\NormalTok{f0}\OperatorTok{,} \OperatorTok{\&}\NormalTok{f1}\OperatorTok{,} \OperatorTok{\&}\NormalTok{f2}\OperatorTok{,} \OperatorTok{\&}\NormalTok{f3}\OperatorTok{,} \OperatorTok{\&}\NormalTok{f4}\OperatorTok{);}
    \OperatorTok{*}\NormalTok{g0 }\OperatorTok{=}\NormalTok{ pow}\OperatorTok{(}\NormalTok{f0}\OperatorTok{,}\NormalTok{ pp}\OperatorTok{);}
    \OperatorTok{*}\NormalTok{g1 }\OperatorTok{=}\NormalTok{ pp }\OperatorTok{*}\NormalTok{ f1 }\OperatorTok{*}\NormalTok{ pow}\OperatorTok{(}\NormalTok{f0}\OperatorTok{,}\NormalTok{ pp }\OperatorTok{{-}} \FloatTok{1.0}\OperatorTok{);}
    \OperatorTok{*}\NormalTok{g2 }\OperatorTok{=}\NormalTok{ pp }\OperatorTok{*} \OperatorTok{(}\NormalTok{pp }\OperatorTok{{-}} \FloatTok{1.0}\OperatorTok{)} \OperatorTok{*}\NormalTok{ pow}\OperatorTok{(}\NormalTok{f0}\OperatorTok{,}\NormalTok{ pp }\OperatorTok{{-}} \FloatTok{2.0}\OperatorTok{)} \OperatorTok{*}\NormalTok{ SQUARE}\OperatorTok{(}\NormalTok{f1}\OperatorTok{);}
    \OperatorTok{*}\NormalTok{g2 }\OperatorTok{+=}\NormalTok{ pp }\OperatorTok{*}\NormalTok{ pow}\OperatorTok{(}\NormalTok{f0}\OperatorTok{,}\NormalTok{ pp }\OperatorTok{{-}} \FloatTok{1.0}\OperatorTok{)} \OperatorTok{*}\NormalTok{ f2}\OperatorTok{;}
    \OperatorTok{*}\NormalTok{g3 }\OperatorTok{=}\NormalTok{ pp }\OperatorTok{*} \OperatorTok{(}\NormalTok{pp }\OperatorTok{{-}} \FloatTok{1.0}\OperatorTok{)} \OperatorTok{*} \OperatorTok{(}\NormalTok{pp }\OperatorTok{{-}} \FloatTok{2.0}\OperatorTok{)} \OperatorTok{*}\NormalTok{ pow}\OperatorTok{(}\NormalTok{f0}\OperatorTok{,}\NormalTok{ pp }\OperatorTok{{-}} \FloatTok{3.0}\OperatorTok{)} \OperatorTok{*}\NormalTok{ THIRD}\OperatorTok{(}\NormalTok{f1}\OperatorTok{);}
    \OperatorTok{*}\NormalTok{g3 }\OperatorTok{+=} \FloatTok{3.0} \OperatorTok{*}\NormalTok{ pp }\OperatorTok{*} \OperatorTok{(}\NormalTok{pp }\OperatorTok{{-}} \FloatTok{1.0}\OperatorTok{)} \OperatorTok{*}\NormalTok{ pow}\OperatorTok{(}\NormalTok{f0}\OperatorTok{,}\NormalTok{ pp }\OperatorTok{{-}} \FloatTok{2.0}\OperatorTok{)} \OperatorTok{*}\NormalTok{ f1 }\OperatorTok{*}\NormalTok{ f2}\OperatorTok{;}
    \OperatorTok{*}\NormalTok{g3 }\OperatorTok{+=}\NormalTok{ pp }\OperatorTok{*}\NormalTok{ pow}\OperatorTok{(}\NormalTok{f0}\OperatorTok{,}\NormalTok{ pp }\OperatorTok{{-}} \FloatTok{1.0}\OperatorTok{)} \OperatorTok{*}\NormalTok{ f3}\OperatorTok{;}
    \OperatorTok{*}\NormalTok{g4 }\OperatorTok{=}\NormalTok{ pp }\OperatorTok{*} \OperatorTok{(}\NormalTok{pp }\OperatorTok{{-}} \FloatTok{1.0}\OperatorTok{)} \OperatorTok{*} \OperatorTok{(}\NormalTok{pp }\OperatorTok{{-}} \FloatTok{2.0}\OperatorTok{)} \OperatorTok{*} \OperatorTok{(}\NormalTok{pp }\OperatorTok{{-}} \FloatTok{3.0}\OperatorTok{)} \OperatorTok{*}\NormalTok{ pow}\OperatorTok{(}\NormalTok{f0}\OperatorTok{,}\NormalTok{ pp }\OperatorTok{{-}} \FloatTok{4.0}\OperatorTok{)} \OperatorTok{*}
\NormalTok{          FOURTH}\OperatorTok{(}\NormalTok{f1}\OperatorTok{);}
    \OperatorTok{*}\NormalTok{g4 }\OperatorTok{+=} \FloatTok{6.0} \OperatorTok{*}\NormalTok{ pp }\OperatorTok{*} \OperatorTok{(}\NormalTok{pp }\OperatorTok{{-}} \FloatTok{1.0}\OperatorTok{)} \OperatorTok{*} \OperatorTok{(}\NormalTok{pp }\OperatorTok{{-}} \FloatTok{2.0}\OperatorTok{)} \OperatorTok{*}\NormalTok{ pow}\OperatorTok{(}\NormalTok{f0}\OperatorTok{,}\NormalTok{ pp }\OperatorTok{{-}} \FloatTok{3.0}\OperatorTok{)} \OperatorTok{*}\NormalTok{ SQUARE}\OperatorTok{(}\NormalTok{f1}\OperatorTok{)} \OperatorTok{*}
\NormalTok{           f2}\OperatorTok{;}
    \OperatorTok{*}\NormalTok{g4 }\OperatorTok{+=} \FloatTok{4.0} \OperatorTok{*}\NormalTok{ pp }\OperatorTok{*} \OperatorTok{(}\NormalTok{pp }\OperatorTok{{-}} \FloatTok{1.0}\OperatorTok{)} \OperatorTok{*}\NormalTok{ pow}\OperatorTok{(}\NormalTok{f0}\OperatorTok{,}\NormalTok{ pp }\OperatorTok{{-}} \FloatTok{2.0}\OperatorTok{)} \OperatorTok{*} \OperatorTok{(}\NormalTok{f1 }\OperatorTok{*}\NormalTok{ f3}\OperatorTok{);}
    \OperatorTok{*}\NormalTok{g4 }\OperatorTok{+=} \FloatTok{3.0} \OperatorTok{*}\NormalTok{ pp }\OperatorTok{*} \OperatorTok{(}\NormalTok{pp }\OperatorTok{{-}} \FloatTok{1.0}\OperatorTok{)} \OperatorTok{*}\NormalTok{ pow}\OperatorTok{(}\NormalTok{f0}\OperatorTok{,}\NormalTok{ pp }\OperatorTok{{-}} \FloatTok{2.0}\OperatorTok{)} \OperatorTok{*}\NormalTok{ SQUARE}\OperatorTok{(}\NormalTok{f2}\OperatorTok{);}
    \OperatorTok{*}\NormalTok{g4 }\OperatorTok{+=}\NormalTok{ pp }\OperatorTok{*}\NormalTok{ pow}\OperatorTok{(}\NormalTok{f0}\OperatorTok{,}\NormalTok{ pp }\OperatorTok{{-}} \FloatTok{1.0}\OperatorTok{)} \OperatorTok{*}\NormalTok{ f4}\OperatorTok{;}
\OperatorTok{\}}

\DataTypeTok{void}\NormalTok{ fStressFaaDiBruno}\OperatorTok{(}\DataTypeTok{const} \DataTypeTok{double} \OperatorTok{*}\NormalTok{x}\OperatorTok{,} \DataTypeTok{const} \DataTypeTok{int} \OperatorTok{*}\NormalTok{n}\OperatorTok{,} \DataTypeTok{const} \DataTypeTok{int} \OperatorTok{*}\NormalTok{p}\OperatorTok{,}
                       \DataTypeTok{const} \DataTypeTok{int} \OperatorTok{*}\NormalTok{u}\OperatorTok{,} \DataTypeTok{const} \DataTypeTok{int} \OperatorTok{*}\NormalTok{v}\OperatorTok{,} \DataTypeTok{const} \DataTypeTok{int} \OperatorTok{*}\NormalTok{fNumber}\OperatorTok{,}
                       \DataTypeTok{const} \DataTypeTok{double} \OperatorTok{*}\NormalTok{par}\OperatorTok{,} \DataTypeTok{double} \OperatorTok{*}\NormalTok{ax}\OperatorTok{,} \DataTypeTok{double} \OperatorTok{*}\NormalTok{gx}\OperatorTok{,} \DataTypeTok{double} \OperatorTok{*}\NormalTok{h0}\OperatorTok{,}
                       \DataTypeTok{double} \OperatorTok{*}\NormalTok{h1}\OperatorTok{,} \DataTypeTok{double} \OperatorTok{*}\NormalTok{h2}\OperatorTok{,} \DataTypeTok{double} \OperatorTok{*}\NormalTok{h3}\OperatorTok{,} \DataTypeTok{double} \OperatorTok{*}\NormalTok{h4}\OperatorTok{)} \OperatorTok{\{}
    \DataTypeTok{double}\NormalTok{ f0}\OperatorTok{,}\NormalTok{ f1}\OperatorTok{,}\NormalTok{ f2}\OperatorTok{,}\NormalTok{ f3}\OperatorTok{,}\NormalTok{ f4}\OperatorTok{;}
    \DataTypeTok{int}\NormalTok{ nn }\OperatorTok{=} \OperatorTok{*}\NormalTok{n}\OperatorTok{,}\NormalTok{ uu }\OperatorTok{=} \OperatorTok{*}\NormalTok{u}\OperatorTok{,}\NormalTok{ vv }\OperatorTok{=} \OperatorTok{*}\NormalTok{v}\OperatorTok{,}\NormalTok{ pp }\OperatorTok{=} \OperatorTok{*}\NormalTok{p}\OperatorTok{,}\NormalTok{ np }\OperatorTok{=}\NormalTok{ nn }\OperatorTok{*}\NormalTok{ pp}\OperatorTok{;}
    \OperatorTok{*}\NormalTok{gx }\OperatorTok{=} \FloatTok{0.0}\OperatorTok{;}
    \ControlFlowTok{for} \OperatorTok{(}\DataTypeTok{int}\NormalTok{ i }\OperatorTok{=} \DecValTok{1}\OperatorTok{;}\NormalTok{ i }\OperatorTok{\textless{}=}\NormalTok{ np}\OperatorTok{;}\NormalTok{ i}\OperatorTok{++)} \OperatorTok{\{}
\NormalTok{        ax}\OperatorTok{[}\NormalTok{VINDEX}\OperatorTok{(}\NormalTok{i}\OperatorTok{)]} \OperatorTok{=} \FloatTok{0.0}\OperatorTok{;}
    \OperatorTok{\}}
    \ControlFlowTok{for} \OperatorTok{(}\DataTypeTok{int}\NormalTok{ s }\OperatorTok{=} \DecValTok{1}\OperatorTok{;}\NormalTok{ s }\OperatorTok{\textless{}=}\NormalTok{ pp}\OperatorTok{;}\NormalTok{ s}\OperatorTok{++)} \OperatorTok{\{}
        \DataTypeTok{double}\NormalTok{ xuv }\OperatorTok{=}\NormalTok{ x}\OperatorTok{[}\NormalTok{MINDEX}\OperatorTok{(}\NormalTok{uu}\OperatorTok{,}\NormalTok{ s}\OperatorTok{,}\NormalTok{ nn}\OperatorTok{)]} \OperatorTok{{-}}\NormalTok{ x}\OperatorTok{[}\NormalTok{MINDEX}\OperatorTok{(}\NormalTok{vv}\OperatorTok{,}\NormalTok{ s}\OperatorTok{,}\NormalTok{ nn}\OperatorTok{)];}
\NormalTok{        ax}\OperatorTok{[}\NormalTok{MINDEX}\OperatorTok{(}\NormalTok{uu}\OperatorTok{,}\NormalTok{ s}\OperatorTok{,}\NormalTok{ nn}\OperatorTok{)]} \OperatorTok{=}\NormalTok{ xuv}\OperatorTok{;}
\NormalTok{        ax}\OperatorTok{[}\NormalTok{MINDEX}\OperatorTok{(}\NormalTok{vv}\OperatorTok{,}\NormalTok{ s}\OperatorTok{,}\NormalTok{ nn}\OperatorTok{)]} \OperatorTok{=} \OperatorTok{{-}}\NormalTok{xuv}\OperatorTok{;}
    \OperatorTok{\}}
    \ControlFlowTok{for} \OperatorTok{(}\DataTypeTok{int}\NormalTok{ i }\OperatorTok{=} \DecValTok{1}\OperatorTok{;}\NormalTok{ i }\OperatorTok{\textless{}=}\NormalTok{ np}\OperatorTok{;}\NormalTok{ i}\OperatorTok{++)} \OperatorTok{\{}
        \OperatorTok{*}\NormalTok{gx }\OperatorTok{+=}\NormalTok{ ax}\OperatorTok{[}\NormalTok{VINDEX}\OperatorTok{(}\NormalTok{i}\OperatorTok{)]} \OperatorTok{*}\NormalTok{ x}\OperatorTok{[}\NormalTok{VINDEX}\OperatorTok{(}\NormalTok{i}\OperatorTok{)];}
    \OperatorTok{\}}
    \OperatorTok{(}\DataTypeTok{void}\OperatorTok{)}\NormalTok{fStressPower}\OperatorTok{(}\NormalTok{gx}\OperatorTok{,}\NormalTok{ fNumber}\OperatorTok{,}\NormalTok{ par}\OperatorTok{,} \OperatorTok{\&}\NormalTok{f0}\OperatorTok{,} \OperatorTok{\&}\NormalTok{f1}\OperatorTok{,} \OperatorTok{\&}\NormalTok{f2}\OperatorTok{,} \OperatorTok{\&}\NormalTok{f3}\OperatorTok{,} \OperatorTok{\&}\NormalTok{f4}\OperatorTok{);}
    \OperatorTok{*}\NormalTok{h0 }\OperatorTok{=}\NormalTok{ f0}\OperatorTok{;}
    \ControlFlowTok{for} \OperatorTok{(}\DataTypeTok{int}\NormalTok{ i }\OperatorTok{=} \DecValTok{1}\OperatorTok{;}\NormalTok{ i }\OperatorTok{\textless{}=}\NormalTok{ np}\OperatorTok{;}\NormalTok{ i}\OperatorTok{++)} \OperatorTok{\{}
\NormalTok{        h1}\OperatorTok{[}\NormalTok{VINDEX}\OperatorTok{(}\NormalTok{i}\OperatorTok{)]} \OperatorTok{=} \FloatTok{2.0} \OperatorTok{*}\NormalTok{ f1 }\OperatorTok{*}\NormalTok{ ax}\OperatorTok{[}\NormalTok{VINDEX}\OperatorTok{(}\NormalTok{i}\OperatorTok{)];}
    \OperatorTok{\}}
    \ControlFlowTok{for} \OperatorTok{(}\DataTypeTok{int}\NormalTok{ i }\OperatorTok{=} \DecValTok{1}\OperatorTok{;}\NormalTok{ i }\OperatorTok{\textless{}=}\NormalTok{ np}\OperatorTok{;}\NormalTok{ i}\OperatorTok{++)} \OperatorTok{\{}
        \ControlFlowTok{for} \OperatorTok{(}\DataTypeTok{int}\NormalTok{ j }\OperatorTok{=} \DecValTok{1}\OperatorTok{;}\NormalTok{ j }\OperatorTok{\textless{}=}\NormalTok{ np}\OperatorTok{;}\NormalTok{ j}\OperatorTok{++)} \OperatorTok{\{}
\NormalTok{            h2}\OperatorTok{[}\NormalTok{MINDEX}\OperatorTok{(}\NormalTok{i}\OperatorTok{,}\NormalTok{ j}\OperatorTok{,}\NormalTok{ np}\OperatorTok{)]} \OperatorTok{=} \FloatTok{2.0} \OperatorTok{*}\NormalTok{ f1 }\OperatorTok{*}\NormalTok{ ASEEK}\OperatorTok{(}\NormalTok{nn}\OperatorTok{,}\NormalTok{ pp}\OperatorTok{,}\NormalTok{ uu}\OperatorTok{,}\NormalTok{ vv}\OperatorTok{,}\NormalTok{ i}\OperatorTok{,}\NormalTok{ j}\OperatorTok{)} \OperatorTok{+}
                                   \FloatTok{4.0} \OperatorTok{*}\NormalTok{ f2 }\OperatorTok{*}\NormalTok{ ax}\OperatorTok{[}\NormalTok{VINDEX}\OperatorTok{(}\NormalTok{i}\OperatorTok{)]} \OperatorTok{*}\NormalTok{ ax}\OperatorTok{[}\NormalTok{VINDEX}\OperatorTok{(}\NormalTok{j}\OperatorTok{)];}
        \OperatorTok{\}}
    \OperatorTok{\}}
    \ControlFlowTok{for} \OperatorTok{(}\DataTypeTok{int}\NormalTok{ i }\OperatorTok{=} \DecValTok{1}\OperatorTok{;}\NormalTok{ i }\OperatorTok{\textless{}=}\NormalTok{ np}\OperatorTok{;}\NormalTok{ i}\OperatorTok{++)} \OperatorTok{\{}
        \ControlFlowTok{for} \OperatorTok{(}\DataTypeTok{int}\NormalTok{ j }\OperatorTok{=} \DecValTok{1}\OperatorTok{;}\NormalTok{ j }\OperatorTok{\textless{}=}\NormalTok{ np}\OperatorTok{;}\NormalTok{ j}\OperatorTok{++)} \OperatorTok{\{}
            \ControlFlowTok{for} \OperatorTok{(}\DataTypeTok{int}\NormalTok{ k }\OperatorTok{=} \DecValTok{1}\OperatorTok{;}\NormalTok{ k }\OperatorTok{\textless{}=}\NormalTok{ np}\OperatorTok{;}\NormalTok{ k}\OperatorTok{++)} \OperatorTok{\{}
\NormalTok{                h3}\OperatorTok{[}\NormalTok{AINDEX}\OperatorTok{(}\NormalTok{i}\OperatorTok{,}\NormalTok{ j}\OperatorTok{,}\NormalTok{ k}\OperatorTok{,}\NormalTok{ np}\OperatorTok{,}\NormalTok{ np}\OperatorTok{)]} \OperatorTok{=}
                    \FloatTok{4.0} \OperatorTok{*}\NormalTok{ f2 }\OperatorTok{*}
                    \OperatorTok{((}\NormalTok{ax}\OperatorTok{[}\NormalTok{VINDEX}\OperatorTok{(}\NormalTok{i}\OperatorTok{)]} \OperatorTok{*}\NormalTok{ ASEEK}\OperatorTok{(}\NormalTok{nn}\OperatorTok{,}\NormalTok{ pp}\OperatorTok{,}\NormalTok{ uu}\OperatorTok{,}\NormalTok{ vv}\OperatorTok{,}\NormalTok{ j}\OperatorTok{,}\NormalTok{ k}\OperatorTok{))} \OperatorTok{+}
                     \OperatorTok{(}\NormalTok{ax}\OperatorTok{[}\NormalTok{VINDEX}\OperatorTok{(}\NormalTok{j}\OperatorTok{)]} \OperatorTok{*}\NormalTok{ ASEEK}\OperatorTok{(}\NormalTok{nn}\OperatorTok{,}\NormalTok{ pp}\OperatorTok{,}\NormalTok{ uu}\OperatorTok{,}\NormalTok{ vv}\OperatorTok{,}\NormalTok{ i}\OperatorTok{,}\NormalTok{ k}\OperatorTok{))} \OperatorTok{+}
                     \OperatorTok{(}\NormalTok{ax}\OperatorTok{[}\NormalTok{VINDEX}\OperatorTok{(}\NormalTok{k}\OperatorTok{)]} \OperatorTok{*}\NormalTok{ ASEEK}\OperatorTok{(}\NormalTok{nn}\OperatorTok{,}\NormalTok{ pp}\OperatorTok{,}\NormalTok{ uu}\OperatorTok{,}\NormalTok{ vv}\OperatorTok{,}\NormalTok{ i}\OperatorTok{,}\NormalTok{ j}\OperatorTok{)));}
\NormalTok{                h3}\OperatorTok{[}\NormalTok{AINDEX}\OperatorTok{(}\NormalTok{i}\OperatorTok{,}\NormalTok{ j}\OperatorTok{,}\NormalTok{ k}\OperatorTok{,}\NormalTok{ nn}\OperatorTok{,}\NormalTok{ nn}\OperatorTok{)]} \OperatorTok{+=}
                    \FloatTok{8.0} \OperatorTok{*}\NormalTok{ f3 }\OperatorTok{*}\NormalTok{ ax}\OperatorTok{[}\NormalTok{VINDEX}\OperatorTok{(}\NormalTok{i}\OperatorTok{)]} \OperatorTok{*}\NormalTok{ ax}\OperatorTok{[}\NormalTok{VINDEX}\OperatorTok{(}\NormalTok{j}\OperatorTok{)]} \OperatorTok{*}\NormalTok{ ax}\OperatorTok{[}\NormalTok{VINDEX}\OperatorTok{(}\NormalTok{k}\OperatorTok{)];}
            \OperatorTok{\}}
        \OperatorTok{\}}
    \OperatorTok{\}}
    \ControlFlowTok{for} \OperatorTok{(}\DataTypeTok{int}\NormalTok{ i }\OperatorTok{=} \DecValTok{1}\OperatorTok{;}\NormalTok{ i }\OperatorTok{\textless{}=}\NormalTok{ np}\OperatorTok{;}\NormalTok{ i}\OperatorTok{++)} \OperatorTok{\{}
        \ControlFlowTok{for} \OperatorTok{(}\DataTypeTok{int}\NormalTok{ j }\OperatorTok{=} \DecValTok{1}\OperatorTok{;}\NormalTok{ j }\OperatorTok{\textless{}=}\NormalTok{ np}\OperatorTok{;}\NormalTok{ j}\OperatorTok{++)} \OperatorTok{\{}
            \ControlFlowTok{for} \OperatorTok{(}\DataTypeTok{int}\NormalTok{ k }\OperatorTok{=} \DecValTok{1}\OperatorTok{;}\NormalTok{ k }\OperatorTok{\textless{}=}\NormalTok{ np}\OperatorTok{;}\NormalTok{ k}\OperatorTok{++)} \OperatorTok{\{}
                \ControlFlowTok{for} \OperatorTok{(}\DataTypeTok{int}\NormalTok{ l }\OperatorTok{=} \DecValTok{1}\OperatorTok{;}\NormalTok{ l }\OperatorTok{\textless{}=}\NormalTok{ np}\OperatorTok{;}\NormalTok{ l}\OperatorTok{++)} \OperatorTok{\{}
\NormalTok{                    h4}\OperatorTok{[}\NormalTok{CINDEX}\OperatorTok{(}\NormalTok{i}\OperatorTok{,}\NormalTok{ j}\OperatorTok{,}\NormalTok{ k}\OperatorTok{,}\NormalTok{ l}\OperatorTok{,}\NormalTok{ np}\OperatorTok{,}\NormalTok{ np}\OperatorTok{,}\NormalTok{ np}\OperatorTok{)]} \OperatorTok{=}
                        \FloatTok{4.0} \OperatorTok{*}\NormalTok{ f2 }\OperatorTok{*}
                        \OperatorTok{(}\NormalTok{ASEEK}\OperatorTok{(}\NormalTok{nn}\OperatorTok{,}\NormalTok{ pp}\OperatorTok{,}\NormalTok{ uu}\OperatorTok{,}\NormalTok{ vv}\OperatorTok{,}\NormalTok{ j}\OperatorTok{,}\NormalTok{ l}\OperatorTok{)} \OperatorTok{*}
\NormalTok{                             ASEEK}\OperatorTok{(}\NormalTok{nn}\OperatorTok{,}\NormalTok{ pp}\OperatorTok{,}\NormalTok{ uu}\OperatorTok{,}\NormalTok{ vv}\OperatorTok{,}\NormalTok{ i}\OperatorTok{,}\NormalTok{ k}\OperatorTok{)} \OperatorTok{+}
\NormalTok{                         ASEEK}\OperatorTok{(}\NormalTok{nn}\OperatorTok{,}\NormalTok{ pp}\OperatorTok{,}\NormalTok{ uu}\OperatorTok{,}\NormalTok{ vv}\OperatorTok{,}\NormalTok{ i}\OperatorTok{,}\NormalTok{ l}\OperatorTok{)} \OperatorTok{*}
\NormalTok{                             ASEEK}\OperatorTok{(}\NormalTok{nn}\OperatorTok{,}\NormalTok{ pp}\OperatorTok{,}\NormalTok{ uu}\OperatorTok{,}\NormalTok{ vv}\OperatorTok{,}\NormalTok{ j}\OperatorTok{,}\NormalTok{ k}\OperatorTok{));}
\NormalTok{                    h4}\OperatorTok{[}\NormalTok{CINDEX}\OperatorTok{(}\NormalTok{i}\OperatorTok{,}\NormalTok{ j}\OperatorTok{,}\NormalTok{ k}\OperatorTok{,}\NormalTok{ l}\OperatorTok{,}\NormalTok{ nn}\OperatorTok{,}\NormalTok{ nn}\OperatorTok{,}\NormalTok{ nn}\OperatorTok{)]} \OperatorTok{+=}
                        \FloatTok{16.0} \OperatorTok{*}\NormalTok{ f4 }\OperatorTok{*}\NormalTok{ ax}\OperatorTok{[}\NormalTok{VINDEX}\OperatorTok{(}\NormalTok{i}\OperatorTok{)]} \OperatorTok{*}\NormalTok{ ax}\OperatorTok{[}\NormalTok{VINDEX}\OperatorTok{(}\NormalTok{j}\OperatorTok{)]} \OperatorTok{*}
\NormalTok{                        ax}\OperatorTok{[}\NormalTok{VINDEX}\OperatorTok{(}\NormalTok{k}\OperatorTok{)]} \OperatorTok{*}\NormalTok{ ax}\OperatorTok{[}\NormalTok{VINDEX}\OperatorTok{(}\NormalTok{l}\OperatorTok{)];}
\NormalTok{                    h4}\OperatorTok{[}\NormalTok{CINDEX}\OperatorTok{(}\NormalTok{i}\OperatorTok{,}\NormalTok{ j}\OperatorTok{,}\NormalTok{ k}\OperatorTok{,}\NormalTok{ l}\OperatorTok{,}\NormalTok{ nn}\OperatorTok{,}\NormalTok{ nn}\OperatorTok{,}\NormalTok{ nn}\OperatorTok{)]} \OperatorTok{+=}
                        \FloatTok{8.0} \OperatorTok{*}\NormalTok{ f3 }\OperatorTok{*}\NormalTok{ ASEEK}\OperatorTok{(}\NormalTok{nn}\OperatorTok{,}\NormalTok{ pp}\OperatorTok{,}\NormalTok{ uu}\OperatorTok{,}\NormalTok{ vv}\OperatorTok{,}\NormalTok{ i}\OperatorTok{,}\NormalTok{ l}\OperatorTok{)} \OperatorTok{*}\NormalTok{ ax}\OperatorTok{[}\NormalTok{VINDEX}\OperatorTok{(}\NormalTok{j}\OperatorTok{)]} \OperatorTok{*}
\NormalTok{                        ax}\OperatorTok{[}\NormalTok{VINDEX}\OperatorTok{(}\NormalTok{k}\OperatorTok{)];}
\NormalTok{                    h4}\OperatorTok{[}\NormalTok{CINDEX}\OperatorTok{(}\NormalTok{i}\OperatorTok{,}\NormalTok{ j}\OperatorTok{,}\NormalTok{ k}\OperatorTok{,}\NormalTok{ l}\OperatorTok{,}\NormalTok{ nn}\OperatorTok{,}\NormalTok{ nn}\OperatorTok{,}\NormalTok{ nn}\OperatorTok{)]} \OperatorTok{+=}
                        \FloatTok{8.0} \OperatorTok{*}\NormalTok{ f3 }\OperatorTok{*}\NormalTok{ ASEEK}\OperatorTok{(}\NormalTok{nn}\OperatorTok{,}\NormalTok{ pp}\OperatorTok{,}\NormalTok{ uu}\OperatorTok{,}\NormalTok{ vv}\OperatorTok{,}\NormalTok{ j}\OperatorTok{,}\NormalTok{ l}\OperatorTok{)} \OperatorTok{*}\NormalTok{ ax}\OperatorTok{[}\NormalTok{VINDEX}\OperatorTok{(}\NormalTok{i}\OperatorTok{)]} \OperatorTok{*}
\NormalTok{                        ax}\OperatorTok{[}\NormalTok{VINDEX}\OperatorTok{(}\NormalTok{k}\OperatorTok{)];}
\NormalTok{                    h4}\OperatorTok{[}\NormalTok{CINDEX}\OperatorTok{(}\NormalTok{i}\OperatorTok{,}\NormalTok{ j}\OperatorTok{,}\NormalTok{ k}\OperatorTok{,}\NormalTok{ l}\OperatorTok{,}\NormalTok{ nn}\OperatorTok{,}\NormalTok{ nn}\OperatorTok{,}\NormalTok{ nn}\OperatorTok{)]} \OperatorTok{+=}
                        \FloatTok{8.0} \OperatorTok{*}\NormalTok{ f3 }\OperatorTok{*}\NormalTok{ ASEEK}\OperatorTok{(}\NormalTok{nn}\OperatorTok{,}\NormalTok{ pp}\OperatorTok{,}\NormalTok{ uu}\OperatorTok{,}\NormalTok{ vv}\OperatorTok{,}\NormalTok{ k}\OperatorTok{,}\NormalTok{ l}\OperatorTok{)} \OperatorTok{*}\NormalTok{ ax}\OperatorTok{[}\NormalTok{VINDEX}\OperatorTok{(}\NormalTok{j}\OperatorTok{)]} \OperatorTok{*}
\NormalTok{                        ax}\OperatorTok{[}\NormalTok{VINDEX}\OperatorTok{(}\NormalTok{j}\OperatorTok{)];}
\NormalTok{                    h4}\OperatorTok{[}\NormalTok{CINDEX}\OperatorTok{(}\NormalTok{i}\OperatorTok{,}\NormalTok{ j}\OperatorTok{,}\NormalTok{ k}\OperatorTok{,}\NormalTok{ l}\OperatorTok{,}\NormalTok{ nn}\OperatorTok{,}\NormalTok{ nn}\OperatorTok{,}\NormalTok{ nn}\OperatorTok{)]} \OperatorTok{+=}
                        \FloatTok{8.0} \OperatorTok{*}\NormalTok{ f3 }\OperatorTok{*}\NormalTok{ ASEEK}\OperatorTok{(}\NormalTok{nn}\OperatorTok{,}\NormalTok{ pp}\OperatorTok{,}\NormalTok{ uu}\OperatorTok{,}\NormalTok{ vv}\OperatorTok{,}\NormalTok{ i}\OperatorTok{,}\NormalTok{ k}\OperatorTok{)} \OperatorTok{*}\NormalTok{ ax}\OperatorTok{[}\NormalTok{VINDEX}\OperatorTok{(}\NormalTok{j}\OperatorTok{)]} \OperatorTok{*}
\NormalTok{                        ax}\OperatorTok{[}\NormalTok{VINDEX}\OperatorTok{(}\NormalTok{l}\OperatorTok{)];}
\NormalTok{                    h4}\OperatorTok{[}\NormalTok{CINDEX}\OperatorTok{(}\NormalTok{i}\OperatorTok{,}\NormalTok{ j}\OperatorTok{,}\NormalTok{ k}\OperatorTok{,}\NormalTok{ l}\OperatorTok{,}\NormalTok{ nn}\OperatorTok{,}\NormalTok{ nn}\OperatorTok{,}\NormalTok{ nn}\OperatorTok{)]} \OperatorTok{+=}
                        \FloatTok{8.0} \OperatorTok{*}\NormalTok{ f3 }\OperatorTok{*}\NormalTok{ ASEEK}\OperatorTok{(}\NormalTok{nn}\OperatorTok{,}\NormalTok{ pp}\OperatorTok{,}\NormalTok{ uu}\OperatorTok{,}\NormalTok{ vv}\OperatorTok{,}\NormalTok{ j}\OperatorTok{,}\NormalTok{ k}\OperatorTok{)} \OperatorTok{*}\NormalTok{ ax}\OperatorTok{[}\NormalTok{VINDEX}\OperatorTok{(}\NormalTok{i}\OperatorTok{)]} \OperatorTok{*}
\NormalTok{                        ax}\OperatorTok{[}\NormalTok{VINDEX}\OperatorTok{(}\NormalTok{l}\OperatorTok{)];}
\NormalTok{                    h4}\OperatorTok{[}\NormalTok{CINDEX}\OperatorTok{(}\NormalTok{i}\OperatorTok{,}\NormalTok{ j}\OperatorTok{,}\NormalTok{ k}\OperatorTok{,}\NormalTok{ l}\OperatorTok{,}\NormalTok{ nn}\OperatorTok{,}\NormalTok{ nn}\OperatorTok{,}\NormalTok{ nn}\OperatorTok{)]} \OperatorTok{+=}
                        \FloatTok{8.0} \OperatorTok{*}\NormalTok{ f3 }\OperatorTok{*}\NormalTok{ ASEEK}\OperatorTok{(}\NormalTok{nn}\OperatorTok{,}\NormalTok{ pp}\OperatorTok{,}\NormalTok{ uu}\OperatorTok{,}\NormalTok{ vv}\OperatorTok{,}\NormalTok{ i}\OperatorTok{,}\NormalTok{ j}\OperatorTok{)} \OperatorTok{*}\NormalTok{ ax}\OperatorTok{[}\NormalTok{VINDEX}\OperatorTok{(}\NormalTok{k}\OperatorTok{)]} \OperatorTok{*}
\NormalTok{                        ax}\OperatorTok{[}\NormalTok{VINDEX}\OperatorTok{(}\NormalTok{l}\OperatorTok{)];}
                \OperatorTok{\}}
            \OperatorTok{\}}
        \OperatorTok{\}}
    \OperatorTok{\}}
    \ControlFlowTok{return}\OperatorTok{;}
\OperatorTok{\}}

\DataTypeTok{void}\NormalTok{ faa\_di\_bruno}\OperatorTok{(}\DataTypeTok{const} \DataTypeTok{double} \OperatorTok{*}\NormalTok{x}\OperatorTok{,} \DataTypeTok{const} \DataTypeTok{int} \OperatorTok{*}\NormalTok{n}\OperatorTok{,} \DataTypeTok{const} \DataTypeTok{int} \OperatorTok{*}\NormalTok{fNumber}\OperatorTok{,}
                  \DataTypeTok{const} \DataTypeTok{double} \OperatorTok{*}\NormalTok{par}\OperatorTok{,} \DataTypeTok{const} \DataTypeTok{double} \OperatorTok{*}\NormalTok{a}\OperatorTok{,} \DataTypeTok{double} \OperatorTok{*}\NormalTok{ax}\OperatorTok{,} \DataTypeTok{double} \OperatorTok{*}\NormalTok{gx}\OperatorTok{,}
                  \DataTypeTok{double} \OperatorTok{*}\NormalTok{h0}\OperatorTok{,} \DataTypeTok{double} \OperatorTok{*}\NormalTok{h1}\OperatorTok{,} \DataTypeTok{double} \OperatorTok{*}\NormalTok{h2}\OperatorTok{,} \DataTypeTok{double} \OperatorTok{*}\NormalTok{h3}\OperatorTok{,} \DataTypeTok{double} \OperatorTok{*}\NormalTok{h4}\OperatorTok{)} \OperatorTok{\{}
    \DataTypeTok{double}\NormalTok{ f0}\OperatorTok{,}\NormalTok{ f1}\OperatorTok{,}\NormalTok{ f2}\OperatorTok{,}\NormalTok{ f3}\OperatorTok{,}\NormalTok{ f4}\OperatorTok{;}
    \DataTypeTok{int}\NormalTok{ nn }\OperatorTok{=} \OperatorTok{*}\NormalTok{n}\OperatorTok{;}
    \OperatorTok{*}\NormalTok{gx }\OperatorTok{=} \FloatTok{0.0}\OperatorTok{;}
    \ControlFlowTok{for} \OperatorTok{(}\DataTypeTok{int}\NormalTok{ i }\OperatorTok{=} \DecValTok{1}\OperatorTok{;}\NormalTok{ i }\OperatorTok{\textless{}=}\NormalTok{ nn}\OperatorTok{;}\NormalTok{ i}\OperatorTok{++)} \OperatorTok{\{}
\NormalTok{        ax}\OperatorTok{[}\NormalTok{VINDEX}\OperatorTok{(}\NormalTok{i}\OperatorTok{)]} \OperatorTok{=} \FloatTok{0.0}\OperatorTok{;}
        \ControlFlowTok{for} \OperatorTok{(}\DataTypeTok{int}\NormalTok{ j }\OperatorTok{=} \DecValTok{1}\OperatorTok{;}\NormalTok{ j }\OperatorTok{\textless{}=}\NormalTok{ nn}\OperatorTok{;}\NormalTok{ j}\OperatorTok{++)} \OperatorTok{\{}
\NormalTok{            ax}\OperatorTok{[}\NormalTok{VINDEX}\OperatorTok{(}\NormalTok{i}\OperatorTok{)]} \OperatorTok{+=}\NormalTok{ a}\OperatorTok{[}\NormalTok{MINDEX}\OperatorTok{(}\NormalTok{i}\OperatorTok{,}\NormalTok{ j}\OperatorTok{,}\NormalTok{ nn}\OperatorTok{)]} \OperatorTok{*}\NormalTok{ x}\OperatorTok{[}\NormalTok{VINDEX}\OperatorTok{(}\NormalTok{j}\OperatorTok{)];}
        \OperatorTok{\}}
        \OperatorTok{*}\NormalTok{gx }\OperatorTok{+=}\NormalTok{ ax}\OperatorTok{[}\NormalTok{VINDEX}\OperatorTok{(}\NormalTok{i}\OperatorTok{)]} \OperatorTok{*}\NormalTok{ x}\OperatorTok{[}\NormalTok{VINDEX}\OperatorTok{(}\NormalTok{i}\OperatorTok{)];}
    \OperatorTok{\}}
    \OperatorTok{(}\DataTypeTok{void}\OperatorTok{)}\NormalTok{fStressPower}\OperatorTok{(}\NormalTok{gx}\OperatorTok{,}\NormalTok{ fNumber}\OperatorTok{,}\NormalTok{ par}\OperatorTok{,} \OperatorTok{\&}\NormalTok{f0}\OperatorTok{,} \OperatorTok{\&}\NormalTok{f1}\OperatorTok{,} \OperatorTok{\&}\NormalTok{f2}\OperatorTok{,} \OperatorTok{\&}\NormalTok{f3}\OperatorTok{,} \OperatorTok{\&}\NormalTok{f4}\OperatorTok{);}
    \OperatorTok{*}\NormalTok{h0 }\OperatorTok{=}\NormalTok{ f0}\OperatorTok{;}
    \ControlFlowTok{for} \OperatorTok{(}\DataTypeTok{int}\NormalTok{ i }\OperatorTok{=} \DecValTok{1}\OperatorTok{;}\NormalTok{ i }\OperatorTok{\textless{}=}\NormalTok{ nn}\OperatorTok{;}\NormalTok{ i}\OperatorTok{++)} \OperatorTok{\{}
\NormalTok{        h1}\OperatorTok{[}\NormalTok{VINDEX}\OperatorTok{(}\NormalTok{i}\OperatorTok{)]} \OperatorTok{=} \FloatTok{2.0} \OperatorTok{*}\NormalTok{ f1 }\OperatorTok{*}\NormalTok{ ax}\OperatorTok{[}\NormalTok{VINDEX}\OperatorTok{(}\NormalTok{i}\OperatorTok{)];}
    \OperatorTok{\}}
    \ControlFlowTok{for} \OperatorTok{(}\DataTypeTok{int}\NormalTok{ i }\OperatorTok{=} \DecValTok{1}\OperatorTok{;}\NormalTok{ i }\OperatorTok{\textless{}=}\NormalTok{ nn}\OperatorTok{;}\NormalTok{ i}\OperatorTok{++)} \OperatorTok{\{}
        \ControlFlowTok{for} \OperatorTok{(}\DataTypeTok{int}\NormalTok{ j }\OperatorTok{=} \DecValTok{1}\OperatorTok{;}\NormalTok{ j }\OperatorTok{\textless{}=}\NormalTok{ nn}\OperatorTok{;}\NormalTok{ j}\OperatorTok{++)} \OperatorTok{\{}
\NormalTok{            h2}\OperatorTok{[}\NormalTok{MINDEX}\OperatorTok{(}\NormalTok{i}\OperatorTok{,}\NormalTok{ j}\OperatorTok{,}\NormalTok{ nn}\OperatorTok{)]} \OperatorTok{=} \FloatTok{2.0} \OperatorTok{*}\NormalTok{ f1 }\OperatorTok{*}\NormalTok{ a}\OperatorTok{[}\NormalTok{MINDEX}\OperatorTok{(}\NormalTok{i}\OperatorTok{,}\NormalTok{ j}\OperatorTok{,}\NormalTok{ nn}\OperatorTok{)]} \OperatorTok{+}
                                   \FloatTok{4.0} \OperatorTok{*}\NormalTok{ f2 }\OperatorTok{*}\NormalTok{ ax}\OperatorTok{[}\NormalTok{VINDEX}\OperatorTok{(}\NormalTok{i}\OperatorTok{)]} \OperatorTok{*}\NormalTok{ ax}\OperatorTok{[}\NormalTok{VINDEX}\OperatorTok{(}\NormalTok{j}\OperatorTok{)];}
        \OperatorTok{\}}
    \OperatorTok{\}}
    \ControlFlowTok{for} \OperatorTok{(}\DataTypeTok{int}\NormalTok{ i }\OperatorTok{=} \DecValTok{1}\OperatorTok{;}\NormalTok{ i }\OperatorTok{\textless{}=}\NormalTok{ nn}\OperatorTok{;}\NormalTok{ i}\OperatorTok{++)} \OperatorTok{\{}
        \ControlFlowTok{for} \OperatorTok{(}\DataTypeTok{int}\NormalTok{ j }\OperatorTok{=} \DecValTok{1}\OperatorTok{;}\NormalTok{ j }\OperatorTok{\textless{}=}\NormalTok{ nn}\OperatorTok{;}\NormalTok{ j}\OperatorTok{++)} \OperatorTok{\{}
            \ControlFlowTok{for} \OperatorTok{(}\DataTypeTok{int}\NormalTok{ k }\OperatorTok{=} \DecValTok{1}\OperatorTok{;}\NormalTok{ k }\OperatorTok{\textless{}=}\NormalTok{ nn}\OperatorTok{;}\NormalTok{ k}\OperatorTok{++)} \OperatorTok{\{}
\NormalTok{                h3}\OperatorTok{[}\NormalTok{AINDEX}\OperatorTok{(}\NormalTok{i}\OperatorTok{,}\NormalTok{ j}\OperatorTok{,}\NormalTok{ k}\OperatorTok{,}\NormalTok{ nn}\OperatorTok{,}\NormalTok{ nn}\OperatorTok{)]} \OperatorTok{=}
                    \FloatTok{4.0} \OperatorTok{*}\NormalTok{ f2 }\OperatorTok{*}
                    \OperatorTok{((}\NormalTok{ax}\OperatorTok{[}\NormalTok{VINDEX}\OperatorTok{(}\NormalTok{i}\OperatorTok{)]} \OperatorTok{*}\NormalTok{ a}\OperatorTok{[}\NormalTok{MINDEX}\OperatorTok{(}\NormalTok{j}\OperatorTok{,}\NormalTok{ k}\OperatorTok{,}\NormalTok{ nn}\OperatorTok{)])} \OperatorTok{+}
                     \OperatorTok{(}\NormalTok{ax}\OperatorTok{[}\NormalTok{VINDEX}\OperatorTok{(}\NormalTok{j}\OperatorTok{)]} \OperatorTok{*}\NormalTok{ a}\OperatorTok{[}\NormalTok{MINDEX}\OperatorTok{(}\NormalTok{i}\OperatorTok{,}\NormalTok{ k}\OperatorTok{,}\NormalTok{ nn}\OperatorTok{)])} \OperatorTok{+}
                     \OperatorTok{(}\NormalTok{ax}\OperatorTok{[}\NormalTok{VINDEX}\OperatorTok{(}\NormalTok{k}\OperatorTok{)]} \OperatorTok{*}\NormalTok{ a}\OperatorTok{[}\NormalTok{MINDEX}\OperatorTok{(}\NormalTok{i}\OperatorTok{,}\NormalTok{ j}\OperatorTok{,}\NormalTok{ nn}\OperatorTok{)]));}
\NormalTok{                h3}\OperatorTok{[}\NormalTok{AINDEX}\OperatorTok{(}\NormalTok{i}\OperatorTok{,}\NormalTok{ j}\OperatorTok{,}\NormalTok{ k}\OperatorTok{,}\NormalTok{ nn}\OperatorTok{,}\NormalTok{ nn}\OperatorTok{)]} \OperatorTok{+=}
                    \FloatTok{8.0} \OperatorTok{*}\NormalTok{ f3 }\OperatorTok{*}\NormalTok{ ax}\OperatorTok{[}\NormalTok{VINDEX}\OperatorTok{(}\NormalTok{i}\OperatorTok{)]} \OperatorTok{*}\NormalTok{ ax}\OperatorTok{[}\NormalTok{VINDEX}\OperatorTok{(}\NormalTok{j}\OperatorTok{)]} \OperatorTok{*}\NormalTok{ ax}\OperatorTok{[}\NormalTok{VINDEX}\OperatorTok{(}\NormalTok{k}\OperatorTok{)];}
            \OperatorTok{\}}
        \OperatorTok{\}}
    \OperatorTok{\}}
    \ControlFlowTok{for} \OperatorTok{(}\DataTypeTok{int}\NormalTok{ i }\OperatorTok{=} \DecValTok{1}\OperatorTok{;}\NormalTok{ i }\OperatorTok{\textless{}=}\NormalTok{ nn}\OperatorTok{;}\NormalTok{ i}\OperatorTok{++)} \OperatorTok{\{}
        \ControlFlowTok{for} \OperatorTok{(}\DataTypeTok{int}\NormalTok{ j }\OperatorTok{=} \DecValTok{1}\OperatorTok{;}\NormalTok{ j }\OperatorTok{\textless{}=}\NormalTok{ nn}\OperatorTok{;}\NormalTok{ j}\OperatorTok{++)} \OperatorTok{\{}
            \ControlFlowTok{for} \OperatorTok{(}\DataTypeTok{int}\NormalTok{ k }\OperatorTok{=} \DecValTok{1}\OperatorTok{;}\NormalTok{ k }\OperatorTok{\textless{}=}\NormalTok{ nn}\OperatorTok{;}\NormalTok{ k}\OperatorTok{++)} \OperatorTok{\{}
                \ControlFlowTok{for} \OperatorTok{(}\DataTypeTok{int}\NormalTok{ l }\OperatorTok{=} \DecValTok{1}\OperatorTok{;}\NormalTok{ l }\OperatorTok{\textless{}=}\NormalTok{ nn}\OperatorTok{;}\NormalTok{ l}\OperatorTok{++)} \OperatorTok{\{}
\NormalTok{                    h4}\OperatorTok{[}\NormalTok{CINDEX}\OperatorTok{(}\NormalTok{i}\OperatorTok{,}\NormalTok{ j}\OperatorTok{,}\NormalTok{ k}\OperatorTok{,}\NormalTok{ l}\OperatorTok{,}\NormalTok{ nn}\OperatorTok{,}\NormalTok{ nn}\OperatorTok{,}\NormalTok{ nn}\OperatorTok{)]} \OperatorTok{=}
                        \FloatTok{4.0} \OperatorTok{*}\NormalTok{ f2 }\OperatorTok{*}
                        \OperatorTok{(}\NormalTok{a}\OperatorTok{[}\NormalTok{MINDEX}\OperatorTok{(}\NormalTok{j}\OperatorTok{,}\NormalTok{ l}\OperatorTok{,}\NormalTok{ nn}\OperatorTok{)]} \OperatorTok{*}\NormalTok{ a}\OperatorTok{[}\NormalTok{MINDEX}\OperatorTok{(}\NormalTok{i}\OperatorTok{,}\NormalTok{ k}\OperatorTok{,}\NormalTok{ nn}\OperatorTok{)]} \OperatorTok{+}
\NormalTok{                         a}\OperatorTok{[}\NormalTok{MINDEX}\OperatorTok{(}\NormalTok{i}\OperatorTok{,}\NormalTok{ l}\OperatorTok{,}\NormalTok{ nn}\OperatorTok{)]} \OperatorTok{*}\NormalTok{ a}\OperatorTok{[}\NormalTok{MINDEX}\OperatorTok{(}\NormalTok{j}\OperatorTok{,}\NormalTok{ k}\OperatorTok{,}\NormalTok{ nn}\OperatorTok{)]);}
\NormalTok{                    h4}\OperatorTok{[}\NormalTok{CINDEX}\OperatorTok{(}\NormalTok{i}\OperatorTok{,}\NormalTok{ j}\OperatorTok{,}\NormalTok{ k}\OperatorTok{,}\NormalTok{ l}\OperatorTok{,}\NormalTok{ nn}\OperatorTok{,}\NormalTok{ nn}\OperatorTok{,}\NormalTok{ nn}\OperatorTok{)]} \OperatorTok{+=}
                        \FloatTok{16.0} \OperatorTok{*}\NormalTok{ f4 }\OperatorTok{*}\NormalTok{ ax}\OperatorTok{[}\NormalTok{VINDEX}\OperatorTok{(}\NormalTok{i}\OperatorTok{)]} \OperatorTok{*}\NormalTok{ ax}\OperatorTok{[}\NormalTok{VINDEX}\OperatorTok{(}\NormalTok{j}\OperatorTok{)]} \OperatorTok{*}
\NormalTok{                        ax}\OperatorTok{[}\NormalTok{VINDEX}\OperatorTok{(}\NormalTok{k}\OperatorTok{)]} \OperatorTok{*}\NormalTok{ ax}\OperatorTok{[}\NormalTok{VINDEX}\OperatorTok{(}\NormalTok{l}\OperatorTok{)];}
\NormalTok{                    h4}\OperatorTok{[}\NormalTok{CINDEX}\OperatorTok{(}\NormalTok{i}\OperatorTok{,}\NormalTok{ j}\OperatorTok{,}\NormalTok{ k}\OperatorTok{,}\NormalTok{ l}\OperatorTok{,}\NormalTok{ nn}\OperatorTok{,}\NormalTok{ nn}\OperatorTok{,}\NormalTok{ nn}\OperatorTok{)]} \OperatorTok{+=}
                        \FloatTok{8.0} \OperatorTok{*}\NormalTok{ f3 }\OperatorTok{*}\NormalTok{ a}\OperatorTok{[}\NormalTok{MINDEX}\OperatorTok{(}\NormalTok{i}\OperatorTok{,}\NormalTok{ l}\OperatorTok{,}\NormalTok{ nn}\OperatorTok{)]} \OperatorTok{*}\NormalTok{ ax}\OperatorTok{[}\NormalTok{VINDEX}\OperatorTok{(}\NormalTok{j}\OperatorTok{)]} \OperatorTok{*}
\NormalTok{                        ax}\OperatorTok{[}\NormalTok{VINDEX}\OperatorTok{(}\NormalTok{k}\OperatorTok{)];}
\NormalTok{                    h4}\OperatorTok{[}\NormalTok{CINDEX}\OperatorTok{(}\NormalTok{i}\OperatorTok{,}\NormalTok{ j}\OperatorTok{,}\NormalTok{ k}\OperatorTok{,}\NormalTok{ l}\OperatorTok{,}\NormalTok{ nn}\OperatorTok{,}\NormalTok{ nn}\OperatorTok{,}\NormalTok{ nn}\OperatorTok{)]} \OperatorTok{+=}
                        \FloatTok{8.0} \OperatorTok{*}\NormalTok{ f3 }\OperatorTok{*}\NormalTok{ a}\OperatorTok{[}\NormalTok{MINDEX}\OperatorTok{(}\NormalTok{j}\OperatorTok{,}\NormalTok{ l}\OperatorTok{,}\NormalTok{ nn}\OperatorTok{)]} \OperatorTok{*}\NormalTok{ ax}\OperatorTok{[}\NormalTok{VINDEX}\OperatorTok{(}\NormalTok{i}\OperatorTok{)]} \OperatorTok{*}
\NormalTok{                        ax}\OperatorTok{[}\NormalTok{VINDEX}\OperatorTok{(}\NormalTok{k}\OperatorTok{)];}
\NormalTok{                    h4}\OperatorTok{[}\NormalTok{CINDEX}\OperatorTok{(}\NormalTok{i}\OperatorTok{,}\NormalTok{ j}\OperatorTok{,}\NormalTok{ k}\OperatorTok{,}\NormalTok{ l}\OperatorTok{,}\NormalTok{ nn}\OperatorTok{,}\NormalTok{ nn}\OperatorTok{,}\NormalTok{ nn}\OperatorTok{)]} \OperatorTok{+=}
                        \FloatTok{8.0} \OperatorTok{*}\NormalTok{ f3 }\OperatorTok{*}\NormalTok{ a}\OperatorTok{[}\NormalTok{MINDEX}\OperatorTok{(}\NormalTok{k}\OperatorTok{,}\NormalTok{ l}\OperatorTok{,}\NormalTok{ nn}\OperatorTok{)]} \OperatorTok{*}\NormalTok{ ax}\OperatorTok{[}\NormalTok{VINDEX}\OperatorTok{(}\NormalTok{j}\OperatorTok{)]} \OperatorTok{*}
\NormalTok{                        ax}\OperatorTok{[}\NormalTok{VINDEX}\OperatorTok{(}\NormalTok{j}\OperatorTok{)];}
\NormalTok{                    h4}\OperatorTok{[}\NormalTok{CINDEX}\OperatorTok{(}\NormalTok{i}\OperatorTok{,}\NormalTok{ j}\OperatorTok{,}\NormalTok{ k}\OperatorTok{,}\NormalTok{ l}\OperatorTok{,}\NormalTok{ nn}\OperatorTok{,}\NormalTok{ nn}\OperatorTok{,}\NormalTok{ nn}\OperatorTok{)]} \OperatorTok{+=}
                        \FloatTok{8.0} \OperatorTok{*}\NormalTok{ f3 }\OperatorTok{*}\NormalTok{ a}\OperatorTok{[}\NormalTok{MINDEX}\OperatorTok{(}\NormalTok{i}\OperatorTok{,}\NormalTok{ k}\OperatorTok{,}\NormalTok{ nn}\OperatorTok{)]} \OperatorTok{*}\NormalTok{ ax}\OperatorTok{[}\NormalTok{VINDEX}\OperatorTok{(}\NormalTok{j}\OperatorTok{)]} \OperatorTok{*}
\NormalTok{                        ax}\OperatorTok{[}\NormalTok{VINDEX}\OperatorTok{(}\NormalTok{l}\OperatorTok{)];}
\NormalTok{                    h4}\OperatorTok{[}\NormalTok{CINDEX}\OperatorTok{(}\NormalTok{i}\OperatorTok{,}\NormalTok{ j}\OperatorTok{,}\NormalTok{ k}\OperatorTok{,}\NormalTok{ l}\OperatorTok{,}\NormalTok{ nn}\OperatorTok{,}\NormalTok{ nn}\OperatorTok{,}\NormalTok{ nn}\OperatorTok{)]} \OperatorTok{+=}
                        \FloatTok{8.0} \OperatorTok{*}\NormalTok{ f3 }\OperatorTok{*}\NormalTok{ a}\OperatorTok{[}\NormalTok{MINDEX}\OperatorTok{(}\NormalTok{j}\OperatorTok{,}\NormalTok{ k}\OperatorTok{,}\NormalTok{ nn}\OperatorTok{)]} \OperatorTok{*}\NormalTok{ ax}\OperatorTok{[}\NormalTok{VINDEX}\OperatorTok{(}\NormalTok{i}\OperatorTok{)]} \OperatorTok{*}
\NormalTok{                        ax}\OperatorTok{[}\NormalTok{VINDEX}\OperatorTok{(}\NormalTok{l}\OperatorTok{)];}
\NormalTok{                    h4}\OperatorTok{[}\NormalTok{CINDEX}\OperatorTok{(}\NormalTok{i}\OperatorTok{,}\NormalTok{ j}\OperatorTok{,}\NormalTok{ k}\OperatorTok{,}\NormalTok{ l}\OperatorTok{,}\NormalTok{ nn}\OperatorTok{,}\NormalTok{ nn}\OperatorTok{,}\NormalTok{ nn}\OperatorTok{)]} \OperatorTok{+=}
                        \FloatTok{8.0} \OperatorTok{*}\NormalTok{ f3 }\OperatorTok{*}\NormalTok{ a}\OperatorTok{[}\NormalTok{MINDEX}\OperatorTok{(}\NormalTok{i}\OperatorTok{,}\NormalTok{ j}\OperatorTok{,}\NormalTok{ nn}\OperatorTok{)]} \OperatorTok{*}\NormalTok{ ax}\OperatorTok{[}\NormalTok{VINDEX}\OperatorTok{(}\NormalTok{k}\OperatorTok{)]} \OperatorTok{*}
\NormalTok{                        ax}\OperatorTok{[}\NormalTok{VINDEX}\OperatorTok{(}\NormalTok{l}\OperatorTok{)];}
                \OperatorTok{\}}
            \OperatorTok{\}}
        \OperatorTok{\}}
    \OperatorTok{\}}
    \ControlFlowTok{return}\OperatorTok{;}
\OperatorTok{\}}

\DataTypeTok{void}\NormalTok{ fStressPartials}\OperatorTok{(}\DataTypeTok{const} \DataTypeTok{double} \OperatorTok{*}\NormalTok{x}\OperatorTok{,} \DataTypeTok{const} \DataTypeTok{double} \OperatorTok{*}\NormalTok{w}\OperatorTok{,} \DataTypeTok{const} \DataTypeTok{double} \OperatorTok{*}\NormalTok{delta}\OperatorTok{,}
                     \DataTypeTok{const} \DataTypeTok{int} \OperatorTok{*}\NormalTok{n}\OperatorTok{,} \DataTypeTok{const} \DataTypeTok{int} \OperatorTok{*}\NormalTok{p}\OperatorTok{,} \DataTypeTok{const} \DataTypeTok{int} \OperatorTok{*}\NormalTok{fNumber}\OperatorTok{,}
                     \DataTypeTok{const} \DataTypeTok{double} \OperatorTok{*}\NormalTok{par}\OperatorTok{,} \DataTypeTok{double} \OperatorTok{*}\NormalTok{stress}\OperatorTok{,} \DataTypeTok{double} \OperatorTok{*}\NormalTok{d}\OperatorTok{,} \DataTypeTok{double} \OperatorTok{*}\NormalTok{fd}\OperatorTok{,}
                     \DataTypeTok{double} \OperatorTok{*}\NormalTok{f1}\OperatorTok{,} \DataTypeTok{double} \OperatorTok{*}\NormalTok{f2}\OperatorTok{,} \DataTypeTok{double} \OperatorTok{*}\NormalTok{f3}\OperatorTok{,} \DataTypeTok{double} \OperatorTok{*}\NormalTok{f4}\OperatorTok{)} \OperatorTok{\{}
    \DataTypeTok{double}\NormalTok{ par2 }\OperatorTok{=} \DecValTok{2} \OperatorTok{*} \OperatorTok{*}\NormalTok{par}\OperatorTok{,}\NormalTok{ nn }\OperatorTok{=} \OperatorTok{*}\NormalTok{n}\OperatorTok{,}\NormalTok{ pp }\OperatorTok{=} \OperatorTok{*}\NormalTok{p}\OperatorTok{,}\NormalTok{ np }\OperatorTok{=}\NormalTok{ nn }\OperatorTok{*}\NormalTok{ pp}\OperatorTok{,}\NormalTok{ gx}\OperatorTok{,}\NormalTok{ hx}\OperatorTok{,}\NormalTok{ h0}\OperatorTok{,}\NormalTok{ g0}\OperatorTok{;}
    \DataTypeTok{double} \OperatorTok{*}\NormalTok{ax }\OperatorTok{=} \OperatorTok{(}\DataTypeTok{double} \OperatorTok{*)}\NormalTok{calloc}\OperatorTok{((}\DataTypeTok{size\_t}\OperatorTok{)}\NormalTok{np}\OperatorTok{,} \KeywordTok{sizeof}\OperatorTok{(}\DataTypeTok{double}\OperatorTok{));}
    \DataTypeTok{double} \OperatorTok{*}\NormalTok{h1 }\OperatorTok{=} \OperatorTok{(}\DataTypeTok{double} \OperatorTok{*)}\NormalTok{calloc}\OperatorTok{((}\DataTypeTok{size\_t}\OperatorTok{)}\NormalTok{np}\OperatorTok{,} \KeywordTok{sizeof}\OperatorTok{(}\DataTypeTok{double}\OperatorTok{));}
    \DataTypeTok{double} \OperatorTok{*}\NormalTok{h2 }\OperatorTok{=} \OperatorTok{(}\DataTypeTok{double} \OperatorTok{*)}\NormalTok{calloc}\OperatorTok{((}\DataTypeTok{size\_t}\OperatorTok{)}\NormalTok{SQUARE}\OperatorTok{(}\NormalTok{np}\OperatorTok{),} \KeywordTok{sizeof}\OperatorTok{(}\DataTypeTok{double}\OperatorTok{));}
    \DataTypeTok{double} \OperatorTok{*}\NormalTok{h3 }\OperatorTok{=} \OperatorTok{(}\DataTypeTok{double} \OperatorTok{*)}\NormalTok{calloc}\OperatorTok{((}\DataTypeTok{size\_t}\OperatorTok{)}\NormalTok{THIRD}\OperatorTok{(}\NormalTok{np}\OperatorTok{),} \KeywordTok{sizeof}\OperatorTok{(}\DataTypeTok{double}\OperatorTok{));}
    \DataTypeTok{double} \OperatorTok{*}\NormalTok{h4 }\OperatorTok{=} \OperatorTok{(}\DataTypeTok{double} \OperatorTok{*)}\NormalTok{calloc}\OperatorTok{((}\DataTypeTok{size\_t}\OperatorTok{)}\NormalTok{FOURTH}\OperatorTok{(}\NormalTok{np}\OperatorTok{),} \KeywordTok{sizeof}\OperatorTok{(}\DataTypeTok{double}\OperatorTok{));}
    \DataTypeTok{double} \OperatorTok{*}\NormalTok{g1 }\OperatorTok{=} \OperatorTok{(}\DataTypeTok{double} \OperatorTok{*)}\NormalTok{calloc}\OperatorTok{((}\DataTypeTok{size\_t}\OperatorTok{)}\NormalTok{np}\OperatorTok{,} \KeywordTok{sizeof}\OperatorTok{(}\DataTypeTok{double}\OperatorTok{));}
    \DataTypeTok{double} \OperatorTok{*}\NormalTok{g2 }\OperatorTok{=} \OperatorTok{(}\DataTypeTok{double} \OperatorTok{*)}\NormalTok{calloc}\OperatorTok{((}\DataTypeTok{size\_t}\OperatorTok{)}\NormalTok{SQUARE}\OperatorTok{(}\NormalTok{np}\OperatorTok{),} \KeywordTok{sizeof}\OperatorTok{(}\DataTypeTok{double}\OperatorTok{));}
    \DataTypeTok{double} \OperatorTok{*}\NormalTok{g3 }\OperatorTok{=} \OperatorTok{(}\DataTypeTok{double} \OperatorTok{*)}\NormalTok{calloc}\OperatorTok{((}\DataTypeTok{size\_t}\OperatorTok{)}\NormalTok{THIRD}\OperatorTok{(}\NormalTok{np}\OperatorTok{),} \KeywordTok{sizeof}\OperatorTok{(}\DataTypeTok{double}\OperatorTok{));}
    \DataTypeTok{double} \OperatorTok{*}\NormalTok{g4 }\OperatorTok{=} \OperatorTok{(}\DataTypeTok{double} \OperatorTok{*)}\NormalTok{calloc}\OperatorTok{((}\DataTypeTok{size\_t}\OperatorTok{)}\NormalTok{FOURTH}\OperatorTok{(}\NormalTok{np}\OperatorTok{),} \KeywordTok{sizeof}\OperatorTok{(}\DataTypeTok{double}\OperatorTok{));}
    \OperatorTok{*}\NormalTok{stress }\OperatorTok{=} \FloatTok{0.0}\OperatorTok{;}
    \ControlFlowTok{for} \OperatorTok{(}\DataTypeTok{int}\NormalTok{ j }\OperatorTok{=} \DecValTok{1}\OperatorTok{;}\NormalTok{ j }\OperatorTok{\textless{}=}\NormalTok{ nn }\OperatorTok{{-}} \DecValTok{1}\OperatorTok{;}\NormalTok{ j}\OperatorTok{++)} \OperatorTok{\{}
        \ControlFlowTok{for} \OperatorTok{(}\DataTypeTok{int}\NormalTok{ i }\OperatorTok{=}\NormalTok{ j }\OperatorTok{+} \DecValTok{1}\OperatorTok{;}\NormalTok{ i }\OperatorTok{\textless{}=}\NormalTok{ nn}\OperatorTok{;}\NormalTok{ i}\OperatorTok{++)} \OperatorTok{\{}
            \DataTypeTok{int}\NormalTok{ k }\OperatorTok{=}\NormalTok{ SINDEX}\OperatorTok{(}\NormalTok{i}\OperatorTok{,}\NormalTok{ j}\OperatorTok{,}\NormalTok{ nn}\OperatorTok{);}
            \OperatorTok{(}\DataTypeTok{void}\OperatorTok{)}\NormalTok{fStressFaaDiBruno}\OperatorTok{(}\NormalTok{x}\OperatorTok{,}\NormalTok{ n}\OperatorTok{,}\NormalTok{ p}\OperatorTok{,} \OperatorTok{\&}\NormalTok{i}\OperatorTok{,} \OperatorTok{\&}\NormalTok{j}\OperatorTok{,}\NormalTok{ fNumber}\OperatorTok{,}\NormalTok{ par}\OperatorTok{,}\NormalTok{ ax}\OperatorTok{,} \OperatorTok{\&}\NormalTok{hx}\OperatorTok{,} \OperatorTok{\&}\NormalTok{h0}\OperatorTok{,}
\NormalTok{                                    h1}\OperatorTok{,}\NormalTok{ h2}\OperatorTok{,}\NormalTok{ h3}\OperatorTok{,}\NormalTok{ h4}\OperatorTok{);}
\NormalTok{            d}\OperatorTok{[}\NormalTok{k}\OperatorTok{]} \OperatorTok{=}\NormalTok{ hx}\OperatorTok{;}
\NormalTok{            fd}\OperatorTok{[}\NormalTok{k}\OperatorTok{]} \OperatorTok{=}\NormalTok{ h0}\OperatorTok{;}
            \OperatorTok{*}\NormalTok{stress }\OperatorTok{+=}\NormalTok{ w}\OperatorTok{[}\NormalTok{k}\OperatorTok{]} \OperatorTok{*}\NormalTok{ SQUARE}\OperatorTok{(}\NormalTok{delta}\OperatorTok{[}\NormalTok{k}\OperatorTok{]} \OperatorTok{{-}}\NormalTok{ fd}\OperatorTok{[}\NormalTok{k}\OperatorTok{]);}
            \OperatorTok{(}\DataTypeTok{void}\OperatorTok{)}\NormalTok{fStressFaaDiBruno}\OperatorTok{(}\NormalTok{x}\OperatorTok{,}\NormalTok{ n}\OperatorTok{,}\NormalTok{ p}\OperatorTok{,} \OperatorTok{\&}\NormalTok{i}\OperatorTok{,} \OperatorTok{\&}\NormalTok{j}\OperatorTok{,}\NormalTok{ fNumber}\OperatorTok{,} \OperatorTok{\&}\NormalTok{par2}\OperatorTok{,}\NormalTok{ ax}\OperatorTok{,} \OperatorTok{\&}\NormalTok{gx}\OperatorTok{,}
                                    \OperatorTok{\&}\NormalTok{g0}\OperatorTok{,}\NormalTok{ g1}\OperatorTok{,}\NormalTok{ g2}\OperatorTok{,}\NormalTok{ g3}\OperatorTok{,}\NormalTok{ g4}\OperatorTok{);}
            \ControlFlowTok{for} \OperatorTok{(}\DataTypeTok{int}\NormalTok{ r }\OperatorTok{=} \DecValTok{1}\OperatorTok{;}\NormalTok{ r }\OperatorTok{\textless{}=}\NormalTok{ np}\OperatorTok{;}\NormalTok{ r}\OperatorTok{++)} \OperatorTok{\{}
                \DataTypeTok{int}\NormalTok{ ind }\OperatorTok{=}\NormalTok{ VINDEX}\OperatorTok{(}\NormalTok{r}\OperatorTok{);}
\NormalTok{                f1}\OperatorTok{[}\NormalTok{ind}\OperatorTok{]} \OperatorTok{+=}\NormalTok{ w}\OperatorTok{[}\NormalTok{k}\OperatorTok{]} \OperatorTok{*} \OperatorTok{(}\FloatTok{0.5} \OperatorTok{*}\NormalTok{ g1}\OperatorTok{[}\NormalTok{ind}\OperatorTok{]} \OperatorTok{{-}}\NormalTok{ delta}\OperatorTok{[}\NormalTok{k}\OperatorTok{]} \OperatorTok{*}\NormalTok{ h1}\OperatorTok{[}\NormalTok{ind}\OperatorTok{]);}
                \ControlFlowTok{for} \OperatorTok{(}\DataTypeTok{int}\NormalTok{ s }\OperatorTok{=} \DecValTok{1}\OperatorTok{;}\NormalTok{ s }\OperatorTok{\textless{}=}\NormalTok{ np}\OperatorTok{;}\NormalTok{ s}\OperatorTok{++)} \OperatorTok{\{}
                    \DataTypeTok{int}\NormalTok{ ind }\OperatorTok{=}\NormalTok{ MINDEX}\OperatorTok{(}\NormalTok{r}\OperatorTok{,}\NormalTok{ s}\OperatorTok{,}\NormalTok{ np}\OperatorTok{);}
\NormalTok{                    f2}\OperatorTok{[}\NormalTok{ind}\OperatorTok{]} \OperatorTok{+=}\NormalTok{ w}\OperatorTok{[}\NormalTok{k}\OperatorTok{]} \OperatorTok{*} \OperatorTok{(}\FloatTok{0.5} \OperatorTok{*}\NormalTok{ g2}\OperatorTok{[}\NormalTok{ind}\OperatorTok{]} \OperatorTok{{-}}\NormalTok{ delta}\OperatorTok{[}\NormalTok{k}\OperatorTok{]} \OperatorTok{*}\NormalTok{ h2}\OperatorTok{[}\NormalTok{ind}\OperatorTok{]);}
                    \ControlFlowTok{for} \OperatorTok{(}\DataTypeTok{int}\NormalTok{ t }\OperatorTok{=} \DecValTok{1}\OperatorTok{;}\NormalTok{ t }\OperatorTok{\textless{}=}\NormalTok{ np}\OperatorTok{;}\NormalTok{ t}\OperatorTok{++)} \OperatorTok{\{}
                        \DataTypeTok{int}\NormalTok{ ind }\OperatorTok{=}\NormalTok{ AINDEX}\OperatorTok{(}\NormalTok{r}\OperatorTok{,}\NormalTok{ s}\OperatorTok{,}\NormalTok{ t}\OperatorTok{,}\NormalTok{ np}\OperatorTok{,}\NormalTok{ np}\OperatorTok{);}
\NormalTok{                        f3}\OperatorTok{[}\NormalTok{ind}\OperatorTok{]} \OperatorTok{+=}\NormalTok{ w}\OperatorTok{[}\NormalTok{k}\OperatorTok{]} \OperatorTok{*} \OperatorTok{(}\FloatTok{0.5} \OperatorTok{*}\NormalTok{ g3}\OperatorTok{[}\NormalTok{ind}\OperatorTok{]} \OperatorTok{{-}}\NormalTok{ delta}\OperatorTok{[}\NormalTok{k}\OperatorTok{]} \OperatorTok{*}\NormalTok{ h3}\OperatorTok{[}\NormalTok{ind}\OperatorTok{]);}
                        \ControlFlowTok{for} \OperatorTok{(}\DataTypeTok{int}\NormalTok{ u }\OperatorTok{=} \DecValTok{1}\OperatorTok{;}\NormalTok{ u }\OperatorTok{\textless{}=}\NormalTok{ np}\OperatorTok{;}\NormalTok{ u}\OperatorTok{++)} \OperatorTok{\{}
                            \DataTypeTok{int}\NormalTok{ ind }\OperatorTok{=}\NormalTok{ CINDEX}\OperatorTok{(}\NormalTok{r}\OperatorTok{,}\NormalTok{ s}\OperatorTok{,}\NormalTok{ t}\OperatorTok{,}\NormalTok{ u}\OperatorTok{,}\NormalTok{ np}\OperatorTok{,}\NormalTok{ np}\OperatorTok{,}\NormalTok{ np}\OperatorTok{);}
\NormalTok{                            f4}\OperatorTok{[}\NormalTok{ind}\OperatorTok{]} \OperatorTok{+=}
\NormalTok{                                w}\OperatorTok{[}\NormalTok{k}\OperatorTok{]} \OperatorTok{*} \OperatorTok{(}\FloatTok{0.5} \OperatorTok{*}\NormalTok{ g4}\OperatorTok{[}\NormalTok{ind}\OperatorTok{]} \OperatorTok{{-}}\NormalTok{ delta}\OperatorTok{[}\NormalTok{k}\OperatorTok{]} \OperatorTok{*}\NormalTok{ h4}\OperatorTok{[}\NormalTok{ind}\OperatorTok{]);}
                        \OperatorTok{\}}
                    \OperatorTok{\}}
                \OperatorTok{\}}
            \OperatorTok{\}}
        \OperatorTok{\}}
    \OperatorTok{\}}
    \OperatorTok{*}\NormalTok{stress }\OperatorTok{/=} \FloatTok{2.0}\OperatorTok{;}
\NormalTok{    free}\OperatorTok{(}\NormalTok{ax}\OperatorTok{);}
\NormalTok{    free}\OperatorTok{(}\NormalTok{h1}\OperatorTok{);}
\NormalTok{    free}\OperatorTok{(}\NormalTok{h2}\OperatorTok{);}
\NormalTok{    free}\OperatorTok{(}\NormalTok{h3}\OperatorTok{);}
\NormalTok{    free}\OperatorTok{(}\NormalTok{h4}\OperatorTok{);}
\NormalTok{    free}\OperatorTok{(}\NormalTok{g1}\OperatorTok{);}
\NormalTok{    free}\OperatorTok{(}\NormalTok{g2}\OperatorTok{);}
\NormalTok{    free}\OperatorTok{(}\NormalTok{g3}\OperatorTok{);}
\NormalTok{    free}\OperatorTok{(}\NormalTok{g4}\OperatorTok{);}
\OperatorTok{\}}
\end{Highlighting}
\end{Shaded}

\section*{References}\label{references}
\addcontentsline{toc}{section}{References}

\phantomsection\label{refs}

\end{document}